\documentclass[twocolumn]{aastex62}

\usepackage{amsmath}

\begin{document}

\title{Constraining Temporal Oscillations of Cosmological Parameters Using Type Ia Supernovae}

\author{Sasha R. Brownsberger}
\affiliation{Department of Physics Harvard University, Cambridge, MA 02138 , USA}

\author{Christopher W. Stubbs}
\affiliation{Department of Physics Harvard University, Cambridge, MA 02138 , USA}
\affiliation{Harvard-Smithsonian Center for Astrophysics, Cambridge, MA 02138, USA}

\author{Daniel M. Scolnic}
\affiliation{Kavli Institute for Cosmological Physics, The University of Chicago, Chicago, IL 60637, USA}

\correspondingauthor{Sasha Brownsberger}
\email{sashabrownsberger@g.harvard.edu} 

\shortauthors{Brownsberger, Stubbs, \& Scolnic 2018}

\nocollaboration
\begin{abstract}
The existing set of type Ia supernovae (SNe Ia) is now sufficient to detect oscillatory deviations from the canonical $\Lambda$CDM cosmology. 
We determine that the Fourier spectrum of the Pantheon data set of spectroscopically well-observed SNe Ia is consistent with the predictions of $\Lambda$CDM.  
We also develop and describe two complementary techniques for using SNe Ia to constrain those alternate cosmological models that predict deviations from $\Lambda$CDM that are oscillatory in conformal time. 
The first technique uses the reduced $\chi^2$ statistic to determine the likelihood that the observed data would result from a given model.  
The second technique uses bootstrap analysis to determine the likelihood that the Fourier spectrum of a proposed model could result from statistical fluctuations around $\Lambda$CDM. 
We constrain three oscillatory alternate cosmological models: one in which the dark energy equation of state parameter oscillates around the canonical value of $w_{\Lambda} = -1$, one in which the energy density of dark energy oscillates around its $\Lambda$CDM value, and one in which gravity derives from a scalar field evolving under an oscillatory coupling.
We further determine that any alternate cosmological model that produces distance modulus residuals with a Fourier amplitude of $\simeq 36$ millimags is strongly ruled out, given the existing data, for frequencies between $\simeq 0.08\ \textrm{Gyr}^ {-1} h_{100}$ and $\simeq 80\ \textrm{Gyr}^ {-1} h_{100}$.   
\end{abstract}

\section{INTRODUCTION}
Over the past several decades, astronomers have leveraged the angular power spectrum of the CMB 
\citep{Bennett93, Jones06, Hinshaw13, Planck16}, Baryon Acoustic Oscillations \citep{Eisenstein05, Percival10, desBourboux17, Slepian17, Bautista17, Beutler17, Ross17, Ata18}, 
measurements of the redshift vs flux relation of type Ia supernovae (SNe Ia) \citep{Riess98, Perlmutter99, Barris04, Hicken09, Kessler09, Sullivan11, Takanashi17, Foley18, Scolnic18}, and other cosmological probes \citep{Zhang08, Batista17, Li18} to gain insight into the structure and dynamics of the observable universe.  These probes have generally been in good agreement with the theoretical predictions of a flat universe dominated today by a cosmological constant ($\Lambda$) and cold dark matter (CDM).  The $\Lambda$CDM cosmology has thus become the canonical cosmological model.

As new cosmological probes are developed and as data sets are improved, astronomers continue to test the consistency of the $\Lambda$CDM cosmology with observations.  
With the recently released Pantheon data set of SNe Ia \citep{Scolnic18}, we can study the expansion history of the universe out to a redshift of $z \simeq 2$ for possible deviations from the predictions of $\Lambda$CDM. 
Numerous efforts \citep{Jassal05, Nesseris07, Santos08, Ferrer09, Busti12, Zhe12, deFelice12, Keresztes15, Trashorras16, Tutusaus17, GomezValent17, Zhao17b, Scolnic18, LHuillier18, Costa18, Yang18, Mehrabi18, Durrive18, Dhawan18, Amirhashchi19}
have applied some version of $\chi^2$ analysis to various data sets of SNe Ia to constrain the gradual evolution of various cosmological parameters away from their canonical values.  However, cosmological models with slowly varying parameters do not encompass the whole family of possible alternative cosmologies.  There are cosmological models, including alternate dark energy (DE) models \citep{Puetzfeld04, Feng06, Jain07, Lazkoz10, Jennings10, Felice12, Zhang18, Wang18}, that predict deviations from $\Lambda$CDM that could be both rapid in evolution and small in overall scale.  Such rapidly changing deviations would be hard to detect using the standard $\chi^2$ test \citep{Maor02}, especially in the context of alternate cosmologies that deviate only gradually from $\Lambda$CDM.  It is worth asking if other analysis methods might prove more apt at detecting such deviations. 

In this work, we leverage the Pantheon SNe Ia data set to search for evidence of deviations from the predictions of $\Lambda$CDM that appear spherically symmetric with respect to an Earth-based observer.  
Oscillatory perturbations, in redshift or in time, are more easily matched to the CMB constraint at the redshift of last scattering than cosmological perturbations that evolve monotonically.  Moreover, even absent any theoretical basis for oscillatory anomalies, our profound ignorance of the nature of DE compels us to consider any well-posed analysis of the available observational data. 

We first measure the temporal Fourier spectrum of observed deviations and look for anomalously large Fourier modes.   
We then consider three alternate cosmological models (ACMs) characterized by oscillatory cosmological parameters.  
We constrain the extent to which fundamental cosmological parameters might vary over redshift ranges as small as about $0.05$ and as large as the full redshift range spanned by the Pantheon data of about $2.3$.  We do so by using both $\chi^2$ and Fourier analyses to search for deviations from the predictions of $\Lambda$CDM that manifest as temporally coherent SNe Ia distance modulus residuals.  We determine that the Fourier spectrum of the Pantheon SNe Ia is consistent with the predictions of the $\Lambda$CDM and that, for some classes of ACMs, the Fourier analysis that we develop provides stronger constraints than the typical $\chi^2$ analysis. 

Our work is organized as follows.  
In Section \ref{sec:data}, we discuss the Pantheon data and define the parameters used in subsequent calculations.  In Section \ref{sec:statMethodFourier}, we describe how to use Fourier analysis of SNe Ia to test the predictions of a cosmological model.  In Section \ref{sec:altModels} we discuss the classes of ACMs and the observable signals that we consider in subsequent calculations.  In Section \ref{sec:constrainingMethods}, we detail the reduced $\chi^2$ and Fourier methods that we use to constrain the considered ACMs.  In Section \ref{sec:resultsFourier}, we describe the consistency of the Fourier spectrum of the Pantheon data set with the predictions of $\Lambda$CDM.  In Section \ref{sec:resultsConstrainingModels}, we discuss the parameter constraints that we find for various oscillatory ACMs.  We conclude in Section \ref{sec:discussion} with some thoughts on the current and future potential for Fourier analysis of SNe Ia to provide insights into the expansion history of the universe.  In Appendices \ref{app:fromWODEs}, \ref{app:fromDEODEs} and \ref{app:fromGODEs}, we detail the computational underpinnings of the considered ACMs.  In Appendix \ref{app:lunarGConstraints}, we discuss additional constraints on one of the ACMs.  

\section{DATA AND METHODOLOGY} \label{sec:dataAndMethod}

\subsection{The Pantheon Data and Cosmological Parameters} \label{sec:data}
Our analysis is based on the Pantheon data set of well-observed SNe Ia detailed by \cite{Scolnic18}.  
These data include contributions from the Pan-STARRS1 Medium Deep Field Survey \citep{Chambers16} 
and from numerous previous observational efforts \citep{Riess98, Perlmutter99, Barris04, Krisciunas05, Jha06, Hicken09, Foley09,  Stritzinger11, Meyers12, Kessler13, Ganeshalingam13, Maguire14, Betoule14, Rodney14, Graur14,  Narayan16, Takanashi17, Sako18}. 
Only SNe with well-observed light curves and precisely known redshifts are included.  Even with such stringent criteria, the data set still comprises $1048$ SNe Ia.  In order to derive a self-consistent dataset, we use the released Pantheon distances when only a single scatter model, the `G10' scatter model, is used to determine distance bias corrections.  

\begin{deluxetable*}{C C C C  C }
 \tablecaption{ Best-Fit $\Lambda$CDM Parameters Reported by the Planck Collaboration}
\tablehead{\colhead{$H_{0,\textrm{can}}\  (\textrm{km s} ^ {-1} \textrm{ Mpc}^{-1})$} &  \colhead{$\Omega_{m, 0, \textrm{can}}$}  & \colhead{$\Omega_{r, 0, \textrm{can}}$}     & \colhead{$\Omega_{\Lambda, 0, \textrm{can}}$} & \colhead{Reference} }
\startdata
  $67.31 \pm 0.96$                                                                     & $0.315 \pm 0.013$                      & $(9.28 \pm0.41) \times 10 ^ {-5}$  & $0.685 \pm 0.013$                             & 1 
\enddata
\tablecomments{The columns are the Planck $\Lambda$CDM present-day values of, from left to right, the Hubble parameter, the normalized energy density of mass, of radiation, and of dark energy.  As \cite{Huang15} discuss, $\Omega_{r, 0, \textrm{can}}$ is calculated from $\Omega_{\Lambda, 0, \textrm{can}}$ and the redshift of matter-radiation equality.} 
 \tablerefs{(1) \cite{Planck16}.} \label{tab:canonParams}
\end{deluxetable*}

The luminosity distance, $d_L$, of an observed SNe Ia is determined from the measured flux, $f$, \citep{Scolnic18} and the absolute SNe Ia luminosity, $L$, by
\begin{equation} \label{eq:dLDef}
d_L 
= \sqrt{\frac{L}{4 \pi f}} \ .
\end{equation}
The signals of observed SNe Ia are parameterized by the distance modulus, $\mu$, defined as 
\begin{equation} \label{eq:muDef}
   \begin{aligned}
   \mu = 25 + 5\textrm{log}_\textrm{10} \Big ( \frac{d_L}{1\textrm{Mpc}}\Big)   \ .
   \end{aligned} 
\end{equation}

In the canonical $\Lambda$CDM cosmology, the canonical Hubble parameter, $H_{\textrm{can}}$, is determined by the values of the normalized, global, present day energy densities of matter, $\Omega_{m,0, \textrm{can}}$, of radiation, $\Omega_{r,0,\textrm{can}}$, and of DE, $\Omega_{\Lambda,0,\textrm{can}}$, by
\begin{equation} \label{eq:canonHofZ}
    \begin{aligned} 
    \frac{H_{\textrm{can}}^2(z)}{H_{0,\textrm{can}}^2} = &(1+z)^ 3 \Omega_{m,0,\textrm{can}} \\
                                                                & + (1+z)^ 4 \Omega_{r,0,\textrm{can}} + \Omega_{\Lambda,0,\textrm{can}} \ ,
     \end{aligned}                                                        
\end{equation}
where $H_{0,\textrm{can}}$ is the present day value of $H_{\textrm{can}}$.  

In such a cosmology, the luminosity distance of any observational target, $d_{L,\textrm{can}}$, is determined from the redshift of the target, $z$, and the evolution of $H_{\textrm{can}}$ by 
\begin{equation} \label{eq:FlatFRW}
d_{L,\textrm{can}}(z) = \frac{c(1+z)}{H_{0,\textrm{can}}} \int_0^z dz' \frac{1}{(H_{\textrm{can}}(z') / H_{0,\textrm{can}} )}  \ ,
\end{equation}
where $c$ is the speed of light.  The distance modulus in the canonical reference cosmology, $\mu_{\textrm{can}}$, is 
\begin{equation} \label{eq:canonMu}
    \begin{aligned}
    \mu_{\textrm{can}} (z) =\  &25 
     + 5\textrm{log}_\textrm{10} \Big (\frac{1} {1 \textrm{Mpc} } \frac{c(1+z)}{H_{0,\textrm{can}}} \times \\ 
     &\ \ \ \  \ \ \ \ \ \ \ \ \ \ \int_0^z dz' \frac{1}{(H_{\textrm{can}}(z') / H_{0,\textrm{can}} )} \Big ) 
\ .
    \end{aligned} 
\end{equation}

In this work we will use the Planck measurements of the CMB anisotropies \citep{Planck16} to define our canonical cosmological parameters.  The relevant parameters are listed in Table \ref{tab:canonParams}.  

\begin{figure*}
\centering
\includegraphics[width=1.0\textwidth]{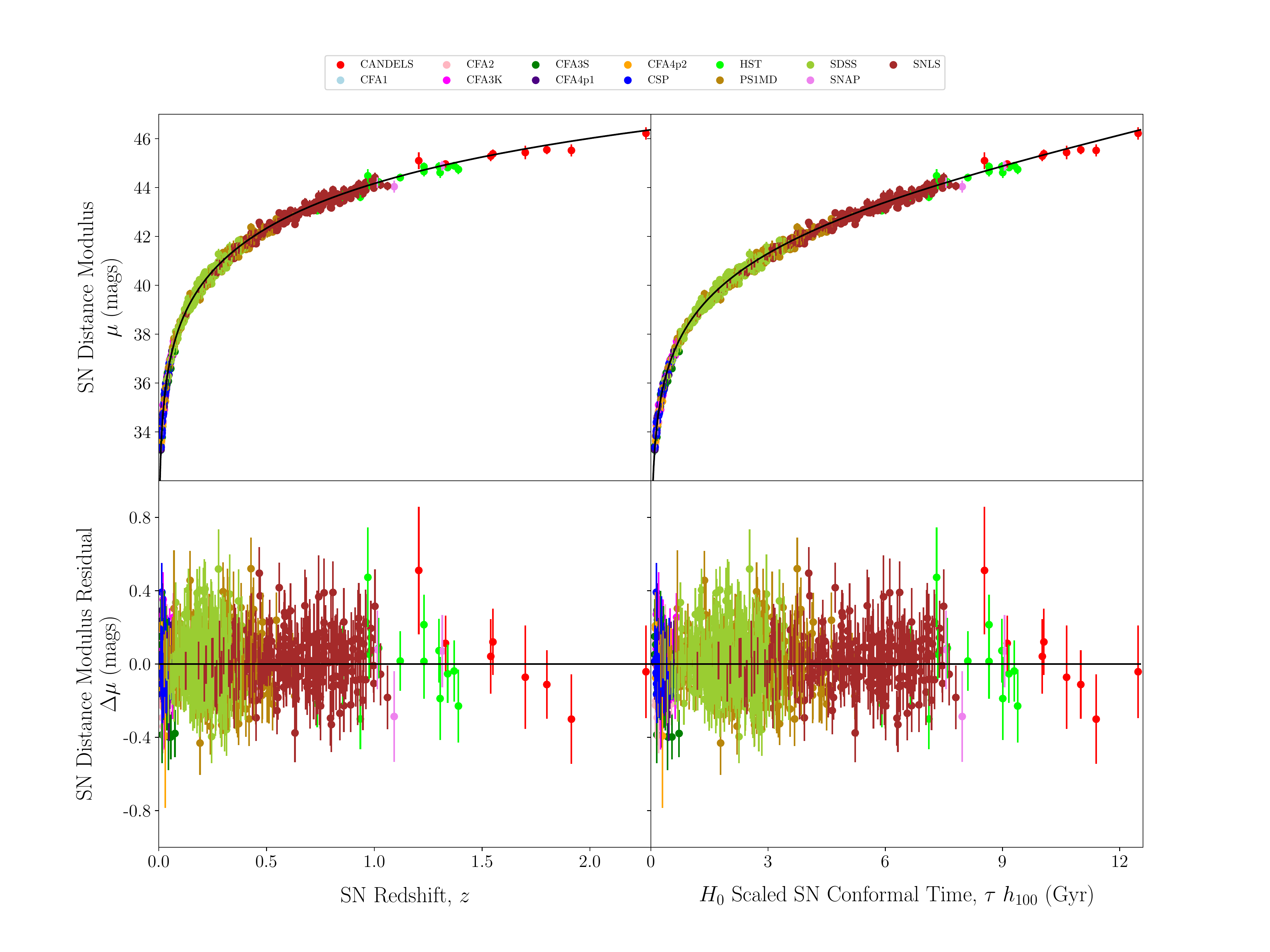}
\caption{\label{fig:muVsZandTau} The distance modulus, $\mu$, and distance modulus residuals, $\Delta \mu$,  vs redshift, $z$, and $H_0$ independent conformal time, $\tau h_{\textrm{100}}$, of the Pantheon data set \citep{Scolnic18} calculated under the Planck $\Lambda$CDM cosmology \citep{Planck16}.  The exact predictions of Planck $\Lambda$CDM are shown as black lines.  The color of each plotted SN indicates the survey that identified the SN, as noted in the legend.  }
\end{figure*}

The observed distance modulus residuals, $\Delta \mu$, are defined as the differences between the observed distance moduli, $\mu$, and the canonical distance moduli, $\mu_{\textrm{can}}$:
\begin{equation} \label{eq:delMuDef}
   \Delta \mu (z) \equiv \mu - \mu_{\textrm{can}}(z) \ .
\end{equation}
If the Planck cosmology incorrectly describes the physics of the expanding universe or if the observable signatures of SNe Ia differ from the expectations of \cite{Scolnic18}, then $\Delta \mu$ will vary in a coherent way in cosmic time.  Conversely, if the canonical cosmology describes the expansion history of the universe with sufficient accuracy and our understanding of SNe Ia physics adequately predicts SNe Ia luminosities, then the measured $\Delta \mu$ values will be consistent with 0, within uncertainties.  

There are many cosmological effects that could alter the expected relation between $\mu$ and $z$, and thus lead to nonzero values of $\Delta \mu$.  Although evidence of direction dependent residuals is certainly an observational signature worth searching for \citep{Gupta08, Campanelli11, Javanmardi15, Jimenez15, Zhao17, Sun18, Deng18, Andrade18}, we consider only effects that alter the SNe signals uniformly in all directions with respect to an Earthly observer.  Under this assumption, the observed flux depends only on the redshift of the SNe Ia:  
\begin{equation}
    \begin{aligned} 
    \Delta \mu(z)&|_{\textrm{spherically symmetric flux}} \\
     &= 5\textrm{log}_\textrm{10} \Big( \frac{H_{0,\textrm{can}} L\sqrt{1/(4 \pi f(z))}}{c (1+z)\int_0^z dz' H_{0,\textrm{can}}/H_{\textrm{can}}(z')} \Big) \ .
    \end{aligned} 
\end{equation}

The true value of $H_0 L$ is not well known, and measurements of $\Delta \mu$ are thus characterized by an unknown, constant offset.  To remove any systematic errors associated with that offset, we subtract the weighted mean residual from the raw $\Delta \mu$, defining the corrected distance modulus residuals by 
\begin{equation} \label{eq:subMean}
\Delta \mu_{i,\textrm{corrected}} = \Delta \mu_{i,\textrm{raw}} - \frac{\sum \Delta \mu_{j,\textrm{raw}} \ \sigma_j ^ {-2} }{  \sum  \sigma_j ^ {-2} }\ ,
\end{equation}
where $\sigma_j$ is the reported uncertainty in the $j^{th}$ SN Ia distance modulus.  Henceforth, $\Delta \mu$ will refer to $\Delta \mu_{\textrm{corrected}}$.  Our measurements are inherently insensitive to constant offsets between observations and the predictions of the $\Lambda$CDM cosmology.  Our work will thus provide no insight into the $H_0$ tension that exists between the most recent Planck observations \citep{Planck16} and other analyses \citep{Riess16, Efstathiou18}.  In this work, we are only interested in and only sensitive to departures from the predictions of the $\Lambda$CDM cosmology that change as the universe evolves.  

Although $z$ is a directly measurable quantity, it is not a natural basis for parameterizing oscillatory modifications to the standard cosmology.  Theoretically-grounded alternate cosmological models are typically tied to the evolution of cosmological parameters in cosmic time.  Moreover, the observable signatures of such models are tied to signals from cosmological sources.  These signals, between their emission and their detection, are warped by cosmic expansion. 

We parameterize our models by the conformal time of an observed signal, 
\begin{equation} \label{eq:tauDef}
\tau \equiv \int_0 ^ z{dz' \frac{1}{H(z')}} \ ,
\end{equation}
in order to correct for the skewing of a cosmological signal due to cosmological expansion.  In particular, global oscillations in some cosmological parameter would manifest, to an observer, as oscillations in conformal time in the observer's reference frame.  Although the mapping between $z$ and $\tau$ does depend on the details of the underlying cosmology, we search only for small perturbations about the canonical cosmology.  The differences between the canonical and alternate $\tau$ values will be small. 
  
Because the Planck value of $H_0$ is highly dependent on the assumed cosmology \citep{Planck16, Bernal16}, ACMs will generically derive a different measurement of $H_0$ from the same Planck data.  When considering alternate cosmological models, we will generally work with the $H_0$ independent quantity, 
\begin{equation}
\tau h_{100} \equiv \tau \frac{H_0}{100 \textrm{km s}^{-1} \textrm{ Mpc} ^ {-1}} \ , 
\end{equation}
from which the values of $\tau$ can be immediately computed for a given $H_0$.  

The measured values of $\mu$ and $\Delta \mu$ are plotted vs $z$ and $\tau h_{100}$ in Figure \ref{fig:muVsZandTau}.  

\subsection{The Fourier Spectrum of SNe Residuals} \label{sec:statMethodFourier}
There are numerous physical phenomena that could lead to a structured relation between $\Delta \mu$ and $\tau$, not all of which can be easily encapsulated in a single family of models.  We utilize a model-agnostic method to assess the consistency of the predictions of $\Lambda$CDM with observations of SNe Ia.  Many previous efforts employ some version of a $\chi ^2$ test to determine if the measured $\Delta \mu$ values could result from statistical fluctuations around the canonical cosmology.  Such tests are not always the most powerful tools for detecting the signatures of certain types of ACMs. 
In this Section, we detail a complementary statistical test based on Fourier analysis of the relation between $\Delta \mu$ and $\tau$.   

The periodogram of $\Delta \mu$ and $\tau$ for $N_{\textrm{SN}}$ observed SNe Ia is defined as 
\begin{equation} \label{eq:periodogramDef}
Q_n \equiv \frac{1}{B} \left|\sum_{j=1}^{N_{\textrm{SN}}} \Delta \mu_j e ^ {-2\pi i f_n \tau_j}\right|^2 \ ,
\end{equation}
and approximates the Fourier power at frequency $f_n$.  The approximate Fourier amplitude at $f_n$, $c_n$, is related to $Q_n$ by 
\begin{equation}\label{eq:fourierPowerToAmp}
c_n = \sqrt{2 Q_n} \ .
\end{equation}
Here, we tune the normalization constant, $B$, such that the largest periodogram value of the function $\Delta \mu(\tau) = A \sin{(2 \pi f \tau)}$ is $A^2/2$ when calculated at the observed $\tau$ values. 

The uneven $\tau$ spacing between observed SNe introduces several complications into the standard Fourier analysis \citep{VenderPlas18}, including a lack of a well-defined Nyquist limit on the maximum measurable frequency, a lack of perfect orthogonality between each of the measured modes, and a difficult-to-characterize noise floor.  We now describe a forward modeling method of Fourier analysis that accounts for these complications.

\subsubsection{Determining the Frequencies to Measure}
First we must determine the Fourier frequencies, $f_n$, at which we will measure $Q_n$.  
For evenly sampled data, the minimum frequency is generally taken to be $1/T$, where $T$ is the total time interval over which the data are measured.  Lower frequencies would lead to a function with insufficient variation over for observations to reliably constrain.  Such a constraint is also appropriate to the case of unevenly sampled data, and we thus adopt a minimum frequency of 
\begin{equation} 
    \begin{aligned} 
        f_{\textrm{min}} & \equiv \frac{h_{100}}{h_{100}(\tau_{\textrm{max}} - \tau_{\textrm{min}})} \\
                                  & \simeq \frac{h_{100} {100\textrm{km s} ^ {-1} \textrm{Mpc} ^ {-1}}}{\int_{z_{\textrm{min} }}^{z_{\textrm{max} }}dz' (H_{0,\textrm{can}} /H_{\textrm{can}}(z'))}  \\
                                  & \simeq 0.08\ \textrm{Gyr}^ {-1} h_{100} \ .
    \end{aligned} 
\end{equation}

If $N$ measured SNe Ia were spaced evenly in time, then a Fourier mode of frequency $f$ in their residuals would produce structure in the periodogram at the aliasing frequencies of $f_{\textrm{alias}, n} = f + n (N/T)$ for all integers $n$ \citep{Meijering02}.  In such a scenario, our analysis would be unable to distinguish between the true structure in the residuals and these aliased modes.  The maximum informative frequency, beyond which measurements would give no additional insight, is the Nyquist frequency, $f_{\textrm{Nyquist}} = N/(2 T)$.  
However, uneven spacing in the data significantly increases the frequency at which aliasing of a true Fourier mode can occur.  In such cases, the maximum informative frequency must be determined on physical grounds.  

If there is some time interval below which the signal of the searched-for phenomenon would be contaminated by other effects, then the maximum frequency can be taken as the inverse of this minimum period.  A number of natural sources of contamination might exist for the Pantheon data set, including:
\begin{itemize}
\item cosmological effects that influence the observed properties of SNe Ia on a particular time scale, 
\item the typical duration of a SNe Ia, 
\item the cadence of observations, 
\item gradual decoherence of an oscillatory signal due to slipping of the phase over time. 
\end{itemize} 

We note that a SN Ia occurring within a deep gravitational well will be characterized by a slightly higher observed redshift and slightly lower observed flux than an equivalent SN Ia occurring at the same moment within a weaker well.  A multitude of similarly sized gravitational wells could introduce deviations from the predicted SNe Ia redshift and flux relation that are of a characteristic scale.  

Galaxy clusters are gravitational wells of a characteristic comoving diameter, 
$D_{\textrm{gc}} \sim 2-10\ \textrm{Mpc}$.  The inverse of the present day light crossing time of a typical galaxy cluster, $c/D_{\textrm{gc}} \simeq 30-140 \ \textrm{Gyr}^{-1}$, could be taken as the maximum frequency that our analysis could constrain.  However, we would like both our minimum and maximum measured frequencies to scale identically with $h_{100}$.  We thus set 
\begin{equation} \label{eq:fMaxDef}
f_{\textrm{max}} = 1000 f_{\textrm{min}} \simeq 80.8 \ \textrm{Gyr} ^ {-1} h_{100} \ ,
\end{equation}
and note that $f_{\textrm{max}}|_{h_{100}=h_{\textrm{Planck}}} = (80.8 \ \textrm{Gyr} ^ {-1}) (0.673) \simeq 54 \ \textrm{Gyr} ^ {-1}$. For the Planck value of $h_{100}$, this definition of $f_{\textrm{max}}$ lies comfortably within the range of maximum periods suggested by the typical size of galaxy clusters.
 If an additional source of signal contamination characterized by a longer time scale comes to light in the future, $f_{\textrm{max}}$ may need to be revised accordingly.  

The $n^{th}$ measured frequency, $f_n$, is the $n^{th}$ multiple of $f_{\textrm{min}}$ with densifying factor $\lambda$, up to $f_{\textrm{max}}$:
\begin{equation}
f_n = {f_{\textrm{min}}  (1 + n/\lambda)}  \ ,
\end{equation}
where $n$ is any non-negative integer such that $f_n < f_{\textrm{max}}$ and where $\lambda$ should be large enough to avoid under sampling the peaks of the periodogram.  We choose $\lambda = 5$, acquiring $N_f = 4995$ measured frequencies, which trace out observable structures in the periodogram (see Figure \ref{fig:trueFourierDecomp}).  

\begin{figure*}
\centering
\includegraphics[width=0.7\textwidth]{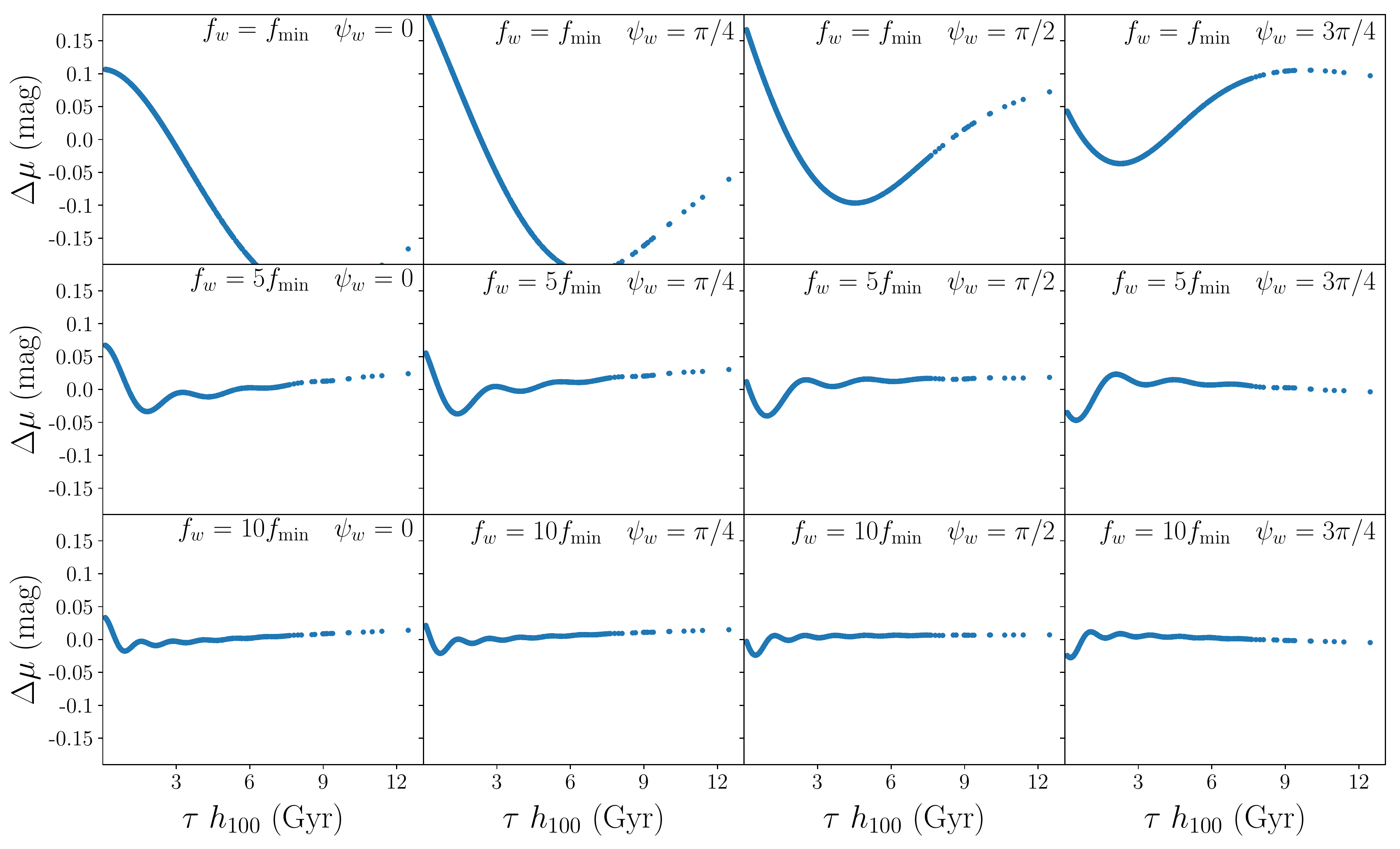}
\caption{\label{fig:muResidsFromWofT} Some examples of theoretical $\Delta \mu$ vs canonical $\tau h_{100}$ values predicted in a cosmology characterized by an alternate DE model with equation of state parameter $w_{\Lambda}(\tau) = -1 + A_{w} \sin{(2 \pi f_{w} \tau + \psi_{w})}$.  Here, $A_w = 1$.  The oscillations of $w_{\Lambda}$ are visible as small oscillations in $\Delta \mu$ that decay as the observed SN grows more distant.  The decaying envelope broadens the Fourier structure of these deviations and weakens the constraining power of the Fourier method described in Section \ref{sec:constrainingMethods}.  We also note that, as the frequency parameter increases, the amplitude of the oscillations diminishes.  The power of SNe Ia to constrain this model diminishes for larger values of $f_{w}$. }
\end{figure*}

\subsubsection{The Canonical Distribution of Periodograms}
With the range of frequencies determined, we now discuss our method for testing the consistency of a periodogram with the canonical cosmology.  

First, we generate a series of artificial data sets.  For the $i^{th}$ observed SN, we determine its conformal time, $\tau_i$, from its measured redshift, $z_i$, and use its reported uncertainty, $\sigma_i$, to determine a random distance modulus residual by drawing from a normal distribution of mean $0$ and of width $\sigma_i$.  Inserting these artificial data into Equation \ref{eq:periodogramDef} furnishes a single realization of the periodogram that could result from observations of the Pantheon SNe Ia in the canonical universe.  

By repeating this process $N_R$ times, we acquire, at each $f_n$, a distribution of periodogram values that could result from a canonical universe given our $\tau_i$ and $\sigma_i$ values.  In this work, $N_R = 1000$.  We now seek to characterize these distinct distributions.  

We define $r_n(Q)$ as the cumulative probability distribution (CPD) of the periodogram value at frequency $f_n$ resulting from a null relation between $\Delta \mu$ and $\tau$.  How `extreme' a measured periodogram value is refers to how unlikely that periodogram value is to occur in the canonical $\Lambda$CDM cosmology, which is to say how close the CPD of that measured periodogram value is to $1$.

 If the $\Delta \mu$ values used in Equation \ref{eq:periodogramDef} were drawn from a function with no dependence on $\tau$, were characterized by Gaussian uncertainties, and were evenly spaced in $\tau$, then $r_n$ would be described by 
\begin{equation} \label{eq:idealCPD}
    \begin{aligned} 
         r_n (Q)  &|_{ \mu(\tau)=0 \textrm{, Gaussian uncertainties, \& evenly spaced } \tau } \\ 
         &= 1 - e^{-Q/c_Q} \ ,
     \end{aligned} 
 \end{equation}
 where $c_Q$ is the mean value of $Q$ \citep{Fisher29}.  
 
Our artificial data are drawn from a null relation between $\Delta \mu$ and $\tau$, are subject to Gaussian uncertainties, and are characterized by the uneven $\tau$ spacing of the observed SNe Ia.  We expect $r_n(Q)$ to be nearly, but not exactly, described by Equation \ref{eq:idealCPD}.  Instead, we fit each $r_n$ with a modified version of the CPD of the generalized normal distribution:
\begin{equation} \label{eq:myGenNormal}
r_n(Q) = \frac{\gamma(1/b_n, (Q/a_n) ^ {b_n})}  {\gamma(1/b_n,\infty)} \ ,
\end{equation}
where $a_n$ and $b_n$ are fit parameters that are determined for each $f_n$, and where $\gamma (x,y) = \int_0^y t^{x-1} e^{-t} dt$ is the lower incomplete gamma function.  Equation \ref{eq:myGenNormal} reduces to Equation \ref{eq:idealCPD} when $b_n = 1$.  By characterizing each $r_n$ with uniquely fitted values of $a_n$ and $b_n$, we provide our algorithm some flexibility in reflecting the unknown structures that could result from the uneven $\tau$ spacing.  We find that the fitted $b_n$ values are generally close to $1$, affirming our expectation that the true CPDs of the periodogram values are nearly described by Equation \ref{eq:idealCPD}.   
We now use our measured best fit CPDs to constrain the periodogram value at each $f_n$ that could be deemed consistent with the canonical cosmology, given the observations.  

All periodograms calculated from Equation \ref{eq:periodogramDef}, whether they are derived from artificial or measured data, consist of a series of peaks of various heights with similar widths, $\Delta f_{\textrm{peak}}$.  We have roughly $N_{\textrm{peak}} = (f_{\textrm{max}} - f_{\textrm{min}})/(\Delta f_{\textrm{peak}}) \simeq 500$ such peaks.  In this work, we regard the height of these peaks as independent random variables.  However, we note that this presumption may be overly simplistic, as the true number of independent peaks when dealing with unevenly sampled data is difficult to precisely quantify \citep{Frescura08}.  

We define the rejection probability, $P_{\textrm{rej}}$, by the statement: `If the likelihood that a periodogram would result from the canonical cosmology is less than $P_{\textrm{rej}}$, then the data from which the periodogram was derived are inconsistent with the canonical cosmology.'
By definition, $r_n (Q)$ is the probability that the periodogram value measured at frequency $f_n$ would be less than $Q$ if the used values of $\Delta \mu$ and $\tau$ resulted from the canonical cosmology.  For a specified $P_{\textrm{rej}}$, the maximum allowable periodogram value at $f_n$, $Q_{\textrm{max},n}$ is given by 
\begin{equation}\label{eq:maxperiodogramDef}
r_n (Q_{\textrm{max},n}) = (1 - P_{\textrm{rej}}) ^ {1/ N_{\textrm{peak}}} \ ,
\end{equation}
where we have accounted for the so called `look-elsewhere' effect \citep{Gross10, Barth18} 
by noting that the likelihood of $N$ independent random variables lying below some probability threshold, $P_{\textrm{thresh}}$, is $(P_{\textrm{thresh}})^{N}$.  
If one or more of the measured $Q_n$ exceed the corresponding $Q_{\textrm{max},n}$, then, for the choice of $P_{\textrm{rej}}$, we reject the hypothesis that the data are described by the canonical cosmology.  Generally, Equation \ref{eq:maxperiodogramDef} is solved numerically for each $Q_{\textrm{max},n}$.  

\subsection{Simulating Observables from Alternative Cosmological Models} \label{sec:altModels}

\begin{figure*} 
\centering
\includegraphics[width=0.7\textwidth]{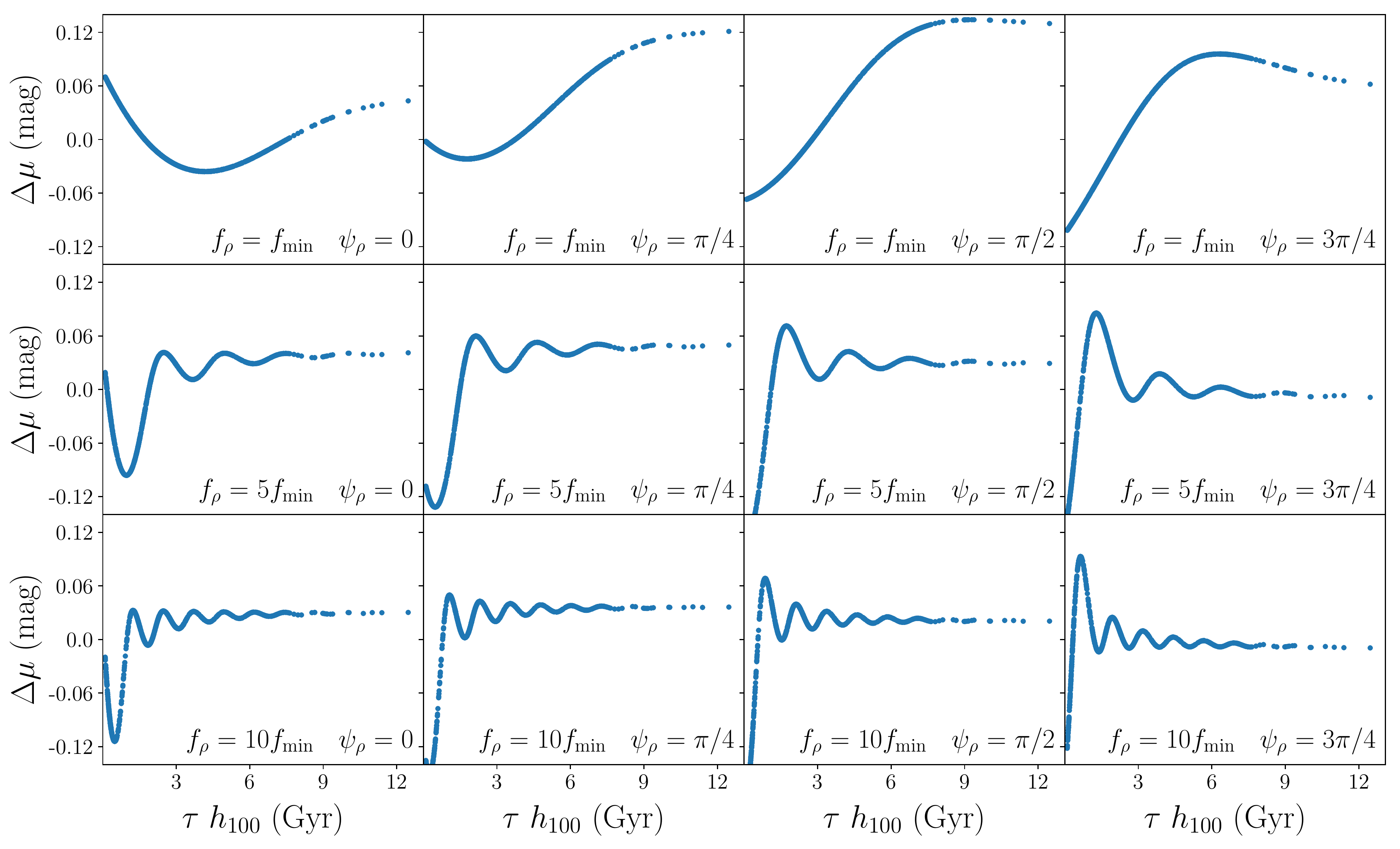}
\caption{\label{fig:muResidsFromDEAofT} Some samples of theoretical $\Delta \mu$ vs canonical $\tau h_{100}$ values predicted in a cosmology with the alternate DE energy density $\rho_{\textrm{DE}}(\tau,z) =  \rho_{ \textrm{DM, can}}(1 + A_{\rho} \sin{(2 \pi f_\rho \tau + \psi_\rho)}) $.  Here, $A_\rho = 0.3$.  As in the case of the ACM with oscillating $w_{\Lambda}$ (Figure \ref{fig:muResidsFromWofT}), the oscillations in $\rho_{\Lambda}$ are visible as oscillations in $\Delta \mu$ and decay as the SNe Ia grow more distant.  However, these oscillations decay less rapidly than in the case of an oscillating $w_{\Lambda}$ because $\Delta \mu$ is related to $\rho_{\Lambda}$ by only one integral, while $\Delta \mu$ and $w_{\Lambda}$ are related by two.  The Periodograms of this model are characterized by two significant peaks: one at the frequency of the model, $f_\rho$, and one at low frequencies resulting from the decaying envelope of the residuals.  } 
\end{figure*}

\begin{figure*} 
\centering
\includegraphics[width=0.7\textwidth]{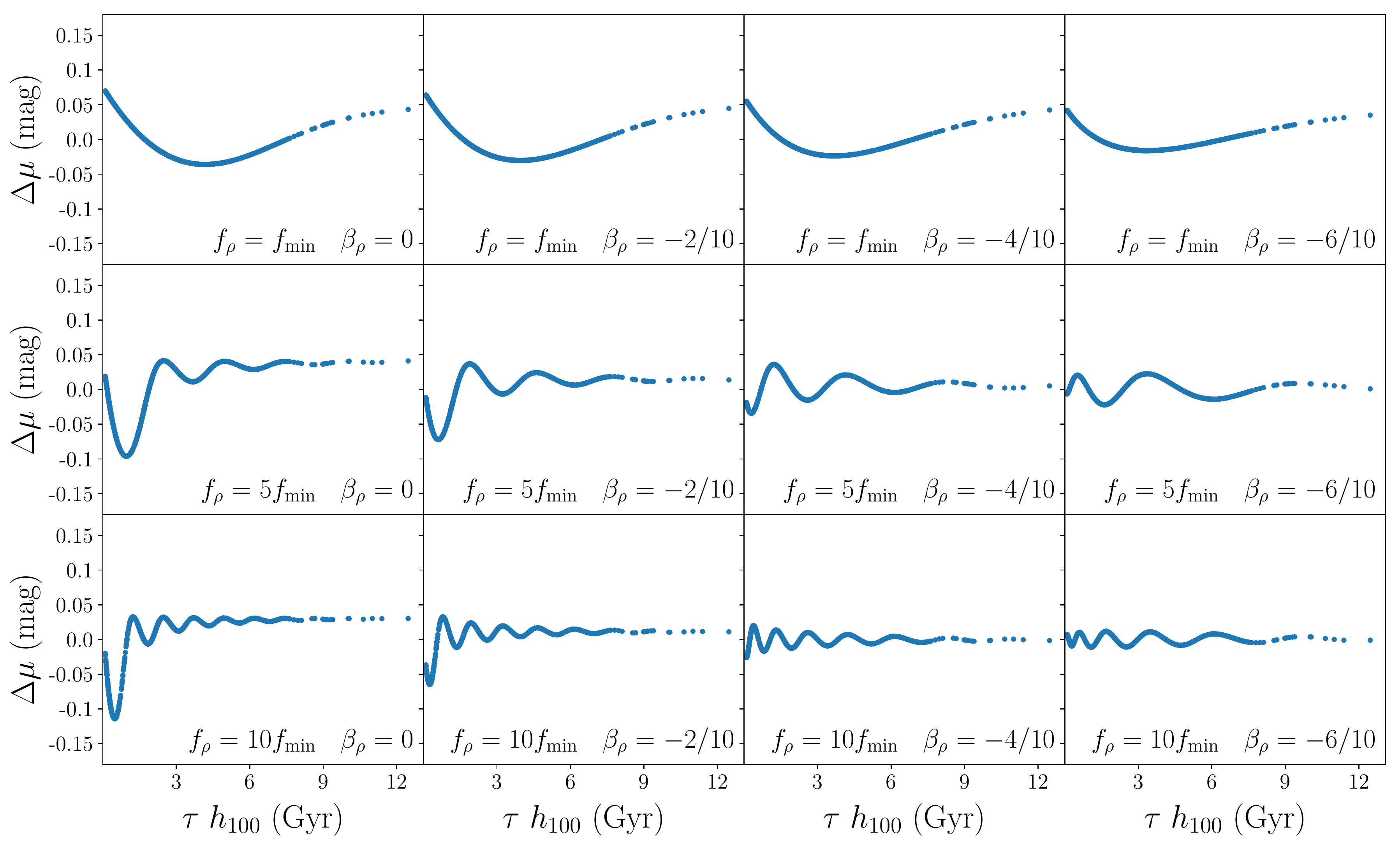}
\caption{\label{fig:muResidsFromDEBofT} Some samples of theoretical $\Delta \mu$ vs canonical $\tau h_{100}$ values predicted in a cosmology with the alternate DE energy density similar to that reported by \cite{Wang18}: $\rho_{\textrm{DE}}(\tau,z) = \rho_{ \textrm{DM, can}} (1 + A_{\rho} (\tau_\rho/\tau)^{\beta_\rho} \sin{(2 \pi f_\rho \tau (\tau/\tau_\rho) ^ {\beta_\rho} + \psi_\rho)}) $.  Here, $A_{\rho}= 0.3$, $\psi_\rho = 0$, and $\tau_\rho \simeq 6.2/ h_{100} \textrm{Gyr}$.  By design, the frequencies of these oscillatory cosmologies decrease in $\tau h_{100}$ when $\beta_\rho < 0$, thus broadening their Fourier peaks.  For some parameter choices, the $\Delta \mu$ vs $\tau h_{100}$ relation is not characterized by a significant decaying envelope.  The Fourier peak at frequency $f_\rho$ dominates the periodogram for those parameter choices. }
\end{figure*}

\begin{figure*} 
\centering
\includegraphics[width=0.7\textwidth]{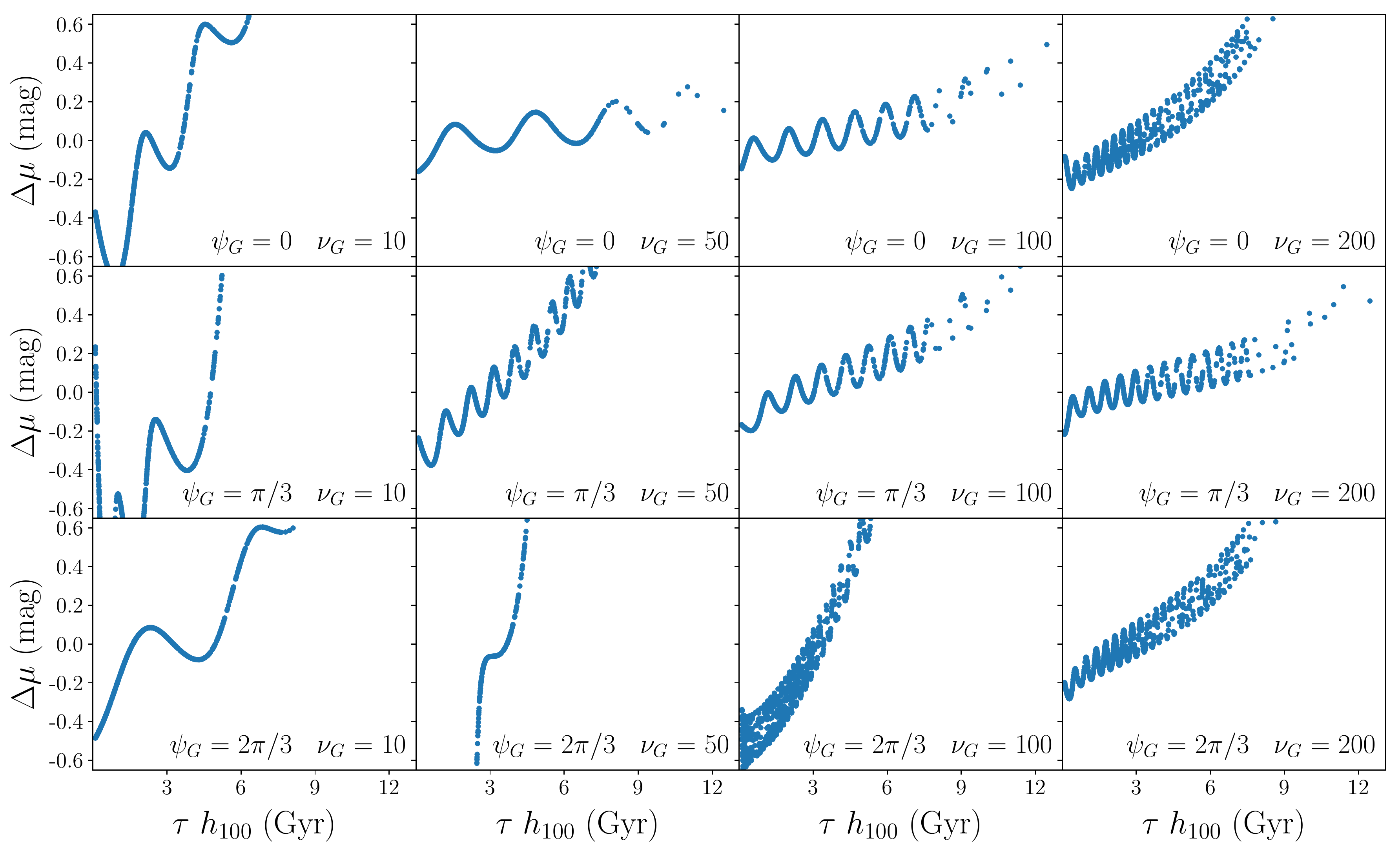}
\caption{\label{fig:muResidsFromGofT} Some samples of theoretical $\Delta \mu$ vs canonical $\tau h_{100}$ values for a cosmology in which gravity derives from a scalar field, $\phi$, that evolves according to the oscillatory coupling $\omega(\phi) = C_G ^{-1} |1 + \epsilon_G - \phi / \phi_0| ^ {-\beta_G}  / (  1 + A_{G}\sin^2{(\nu_G \phi/\phi_0 + \psi_G)})$.  Here, $C_G = 0.02$, $A_G = 7$, $\beta_G = 0.5$, $\epsilon_G$ is tuned to minimize the contemporary value of $dG/dt$, and $\phi_0$ is tuned so that $G$ takes on its contemporary value.  As discussed in Appendix \ref{app:fromGODEs} and by \cite{Gaztanaga01}, the luminosity of SNe Ia depends on the local gravitational constant via a power law relation, and the changing value of $G$ predominantly determines the evolution of $\Delta \mu$.  Because the oscillations in this model are tied to the value of $\phi$, rather than directly to $\tau$, and because $\phi$ evolves according to a nontrivial set of ODEs (Equation \ref{eq:fromGODEs}), the evolution of the cosmological observables is highly sensitive to the particular choice of model parameters.}
\end{figure*} 

We now turn to the task of constraining specific deviations from the predictions of the canonical cosmology.  
In Appendices \ref{app:fromWODEs}, \ref{app:fromDEODEs}, and \ref{app:fromGODEs}, we derive three sets of coupled, first order ODEs (Equations  \ref{eq:fromWODEs},  \ref{eq:fromDEODEs}, and \ref{eq:fromGODEs}) that characterize the evolution of the universe under each of three types of ACMs. 
Once the evolving parameter of interest is specified, these ODEs can be numerically solved to determine the values of $d_L$ predicted by the considered ACM at each measured $\tau$.  

The parameterizations considered in this work are used primarily as phenomenological examples of how Fourier analysis of SNe Ia residuals can be used to constrain ACMs with oscillatory behaviors.  Although these parameterizations are not fundamentally grounded on first principle derivations, generally similar theoretical models are considered in the literature.  Further, versions of such models may prove either to be theoretically viable themselves or to be good approximations of physically sound theories.  Our analysis demonstrates how future authors can and should use Fourier analysis to constrain any ACM characterized by temporal oscillations.  

We first consider an ACM characterized by an evolving DE equation of state (EoS) parameter: $w_{\Lambda} \rightarrow w_{\Lambda}(\tau)$.  Since the existence of DE in the universe was firmly established \citep{Riess98, Perlmutter99}, many fundamental and phenomenological models that propose a DE EoS parameter that deviates from the canonical value of $w_{\Lambda} = -1$ have been considered \citep{Efstathiou99, Cooray99, Chevallier01, Goliath01, Weller02,  Linder03, Puetzfeld04, Feng06, Jain07, Lazkoz10, Jennings10, Felice12, Zhang2015,  Pantazis16}.  
As numerous previous efforts have demonstrated \citep{Jassal05, Barenboim06, Nesseris07, Santos08, Ferrer09, Busti12, Zhe12, deFelice12, Peiris13, Keresztes15, Tutusaus17, GomezValent17, Zhao17b, Scolnic18, LHuillier18, Costa18, Yang18, Davari18, Dhawan18, Amirhashchi19}, the observable properties of SNe Ia are effective tools for constraining these diverse DE models.  

We consider an oscillatory parameterization of $w_{\Lambda}$ similar to those presented by \cite{Xia05, Feng06, Jain07, Liu09, Lazkoz10}:
\begin{equation} \label{eq:wOfTDef}
w_{\Lambda}(\tau) = -1 + A_{w} \sin{(2 \pi f_{w} \tau + \psi_{w})} \ ,
\end{equation}
where $A_{w}$ (unitless), $f_{w}$ (units of 1/[time]), and $\psi_{w}$ (unitless) are constrainable parameters of the model.  

Developing a fundamental field framework for a DE EoS parameter described by Equation \ref{eq:wOfTDef} lies outside the scope of this work.  
 This model does periodically send $w_{\Lambda}$ into the phantom DE region of $w_{\Lambda}<-1$, and a constant value of $w_{\Lambda}<-1$ is difficult to mould into a theoretically sound field theory \citep{Carroll03, Cline04, Hsu04, Sbisa14, Leyva17}.  However, the challenge of reconciling a purely phantom DE model with a fundamental field theory does not render irrelevant any effort to search for the observable signature of an ACM characterized by Equation \ref{eq:wOfTDef}.  
 
Efforts to reconcile a purely phantom DE with a theoretically sound theory persist \citep{Nesseris07, Zhang08, Nunes15, Ludwick17, Albarran17, Ludwick18}, and a value of $w_{\Lambda} < -1$ has not been determined to be wholly inviable.  Further, our analysis assumes that $w_{\Lambda}$ conforms to Equation \ref{eq:wOfTDef} only over the redshift range spanned by the Pantheon data set shown in Figure \ref{fig:muVsZandTau}.  Theoretical problems with phantom DE deriving from the behavior of such DE in the early or future universe are not strictly inconsistent with $w_{\Lambda}$ conforming to Equation \ref{eq:wOfTDef} for the epoch relevant to our analysis.  Finally, and most importantly, we reiterate that the ACMs considered in this work are used primarily as illustrative examples of how Fourier analysis of SNe Ia residuals can be used to constrain broadly similar ACMs.   

In Figure \ref{fig:muResidsFromWofT}, we plot the values of $\Delta \mu$ predicted by the ODEs of Equation \ref{eq:fromWODEs} applied to the specific ACM of Equation \ref{eq:wOfTDef} for various values of  $A_{w}$, $\psi_{w}$, and $f_{w}$.  In Figure \ref{fig:fromWRChiSqr}, we show the constraints that the analyses of Section \ref{sec:constrainingMethods} place on these parameters.  

We next consider an ACM characterized by an alternate DE energy density that evolves according to $\rho_{ \textrm{DM, can}}(z) \rightarrow X(\tau,z) \rho_{ \textrm{DM, can}}(z)$, where $X(\tau,z)$ is some evolving scale factor and $\rho_{ \textrm{DM, can}}(z)$ is the canonical form of the DE energy density. 

In this work, we consider two DE energy density scalings that are oscillatory in $\tau$:  
\begin{equation}  \label{eq:rhoDEAOfTDef}
   \begin{aligned} 
& \rho_{\textrm{DE}}( \tau) = \rho_{ \textrm{DM, can}} (1 + A_{\rho} \sin{(2 \pi f_{\rho} \tau + \psi_{\rho})}) \ ,
   \end{aligned} 
\end{equation}
and
\begin{equation}  \label{eq:rhoDEBOfTDef}
   \begin{aligned} 
\rho_{\textrm{DE}} &(\tau)   =  \rho_{ \textrm{DM, can}}\  \times \\ 
      &   \Big ( 1 + A_{\rho} \Big (\frac{\tau_\rho}{\tau} \Big ) ^ {\beta_\rho} 
           \sin{( 2 \pi f_{\rho} \tau \Big (\frac{\tau}{\tau_\rho} \Big ) ^ {\beta_\rho} + \psi_{\rho})} \Big )  \ , 
   \end{aligned}
\end{equation}
where $\rho_{ \textrm{DM, can}} = (3 c ^2 H_0^2) / (8 \pi G) \Omega_{\Lambda, 0} $ is defined in Equation \ref{eq:canonHofZ}, and where $A_{\rho}$ (unitless), $f_{\rho}$ (units of 1/[time]), $\psi_{\rho}$ (unitless), and $\beta_{\rho}$ (unitless) are constrainable parameters of the model.  Although $\tau_\rho$ (units of [time]) could be removed by redefining other parameters in Equation \ref{eq:rhoDEBOfTDef}, we choose to explicitly include it because it cleanly parameterizes the function's decay timescale.  For Equation \ref{eq:rhoDEBOfTDef}, we fix $\psi_\rho=0$ and $\tau_\rho$ to the middle of the $\tau$ window spanned by the Pantheon data set (see Figure \ref{fig:muVsZandTau}):  $\tau_\rho \simeq 6.2/ h_{100} \textrm{Gyr}$.

We regard the simple oscillatory form of $\rho_{\textrm{DE}}$ described in Equation \ref{eq:rhoDEAOfTDef} as a baseline of the hypothetical signature of oscillating DE.  We use the decaying oscillatory $\rho_{\textrm{DE}}$ of Equation \ref{eq:rhoDEBOfTDef} to replicate the semi-oscillatory $\rho_{\textrm{DE}}$ behavior that \cite{Wang18} claim is a better match to the available cosmological data than the canonical $\Lambda$CDM cosmology.  In Figures \ref{fig:muResidsFromDEAofT} and \ref{fig:muResidsFromDEBofT}, we plot the values of $\Delta \mu$ predicted by the ODEs of Equation \ref{eq:fromDEODEs} given the particular ACMs of Equations \ref{eq:rhoDEAOfTDef} and \ref{eq:rhoDEBOfTDef}  for various values of $f_{\rho}$, $\psi_{\rho}$ and $\beta_{\rho}$.  We show the constraints that the analyses of Section \ref{sec:constrainingMethods} place on these models in Figures \ref{fig:fromDEAFourierLims} and \ref{fig:fromDEBFourierLims}.

The third ACM that we consider is one in which the strength of gravity, $G(\phi)$, is a function of some scalar field, $\phi$, that evolves under some variable coupling, $\omega(\phi)$ (see Appendix \ref{app:fromGODEs}).  Authors have postulated on both phenomenological and theoretical grounds that various constants of nature, including $G$, might evolve slowly in time \citep{Milne37Book, Milne37Article, Dirac37, Brans61, Ivashchuk88, Barrow99, Davidson05, Melnikov09, Iorio16, Kofinas16, Roy17, Tiwari17}.  Several experimental efforts have placed bounds on the recent variation of $G$ \citep{Muller07} and several authors have also used various cosmological signatures (including previous SNe Ia data sets) to constrain the  evolution of $G$ in recent cosmic history \citep{Gaztanaga01, GarciaBerro06, Tomaschitz10, Arun18, Kazantzidis18}.  

In this work, we consider an oscillatory extension of one of the parameterizations of the scalar field coupling considered by \cite{Barrow97}: 
\begin{equation}  \label{eq:GOfTDef}
   \begin{aligned} 
\omega(\phi) = & \frac{1}{C_G} |1 + \epsilon_G - \frac{\phi}{ \phi_0} |^ {-\beta_G} \\  
&  \times \frac{1} {1 + A_G \sin{}^2(\nu_G G_0 \phi/\phi_0 + \psi_G ) } \ , 
   \end{aligned}
\end{equation}
where we have added an inverse, oscillatory scaling that does not produce divergences, where $\epsilon_G$ parameterizes the difference between the current value of $\phi$ and its asymptotic value, where $G_0$ is our contemporary measurement of $G$, where $\phi_0$ is the present day value of $\phi$, and where $C_G$, $\beta_G$, $\nu_G$, $A_G$, and $\psi_G$ (all unitless) are constrainable parameters of the model.  

In this work, we set $A_G = 7$, we set $\beta_G = 0.5$, we tune $\epsilon_G$ so as to minimize the contemporary value of $|dG/dt|$, and we tune $\phi_0$ such that $G(\phi_0) = G_0$.  We recalculate $\epsilon_G$ and $\phi_0$ for every unique set of the constrainable parameter values.  In Figure \ref{fig:muResidsFromGofT}, we plot the values of $\Delta \mu$ predicted by the ODEs of Equation \ref{eq:fromGODEs} given the particular ACM of Equation \ref{eq:GOfTDef} for various values of $\nu_G$ and $\psi_G$.  

In Appendix \ref{app:lunarGConstraints}, we discuss the constraints that the Lunar Ranging Experiment \citep{Muller07} places on Equation \ref{eq:GOfTDef}.  We show these constraints in parallel with the constraints acquired by our analysis in Figure \ref{fig:fromGFourierLims}.   

\subsection{Constraining Alternate Cosmological Models} \label{sec:constrainingMethods}
Here we describe two techniques for determining the portion of an ACM's parameter space that is consistent with the data.  For both constraining methods, we are interested in the consistency of a given ACM with observations relative to the consistency of the $\Lambda$CDM cosmology with those same observations.  When using a particular test statistic to determine the consistency of an ACM with the data, we scale the raw probability of that test statistic by the probability of the same statistic applied to the appropriate null hypothesis.  We then deem a particular ACM consistent or inconsistent with the data based on the value of this probability ratio and some chosen rejection ratio, $R_{\textrm{rej}}$.  If, for example, we choose $R_{\textrm{rej}}$ to be $0.01$, then we reject all ACMs that are $\geq 100$ times more excursive, which is to say less likely, than the corresponding null hypothesis. 

\begin{figure*}
\centering
\includegraphics[width=1.0\textwidth]{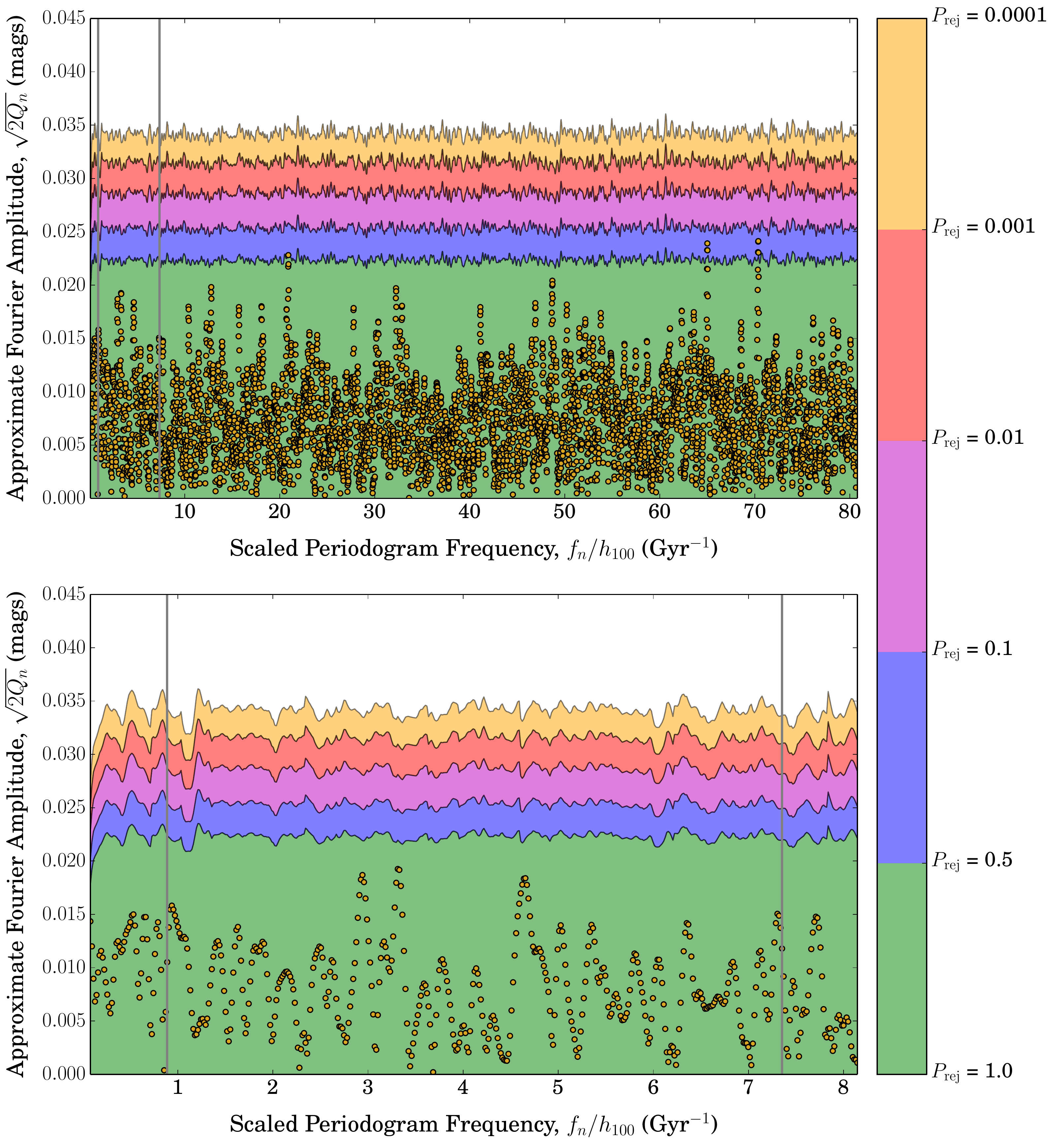}
\caption{\label{fig:trueFourierDecomp} The Fourier decomposition of the Pantheon SN data (dots) compared to the distribution of Fourier decompositions derived from repeated resamplings of artificial null data (shaded regions).  The top plot shows all measured frequencies and the bottom plot shows only the first $500$ measured frequencies.  The gray lines encompass the same frequency range.  The contours indicate the rejection regions for various values of $P_{\textrm{rej}}$.  In the panel depicting all measured Fourier amplitudes, we applied a rectangular smoothing of width $10  f_{\textrm{min}}$ to the threshold Fourier amplitudes.  The most excursive (i.e. least likely) Fourier peak of the Pantheon data set corresponds to $P_{\textrm{rej}} \simeq 0.28$.  Therefore, there is a roughly $28\%$ chance that SNe Ia observations with the redshifts and uncertainties of the Pantheon data taken in the Planck $\Lambda$CDM cosmology would contain at least one Fourier peak more extreme than the most extreme Fourier peak that is actually observed in the Fourier spectrum of the Pantheon data.}
\end{figure*} 

\begin{figure*}
\centering
\includegraphics[width=1.0\textwidth]{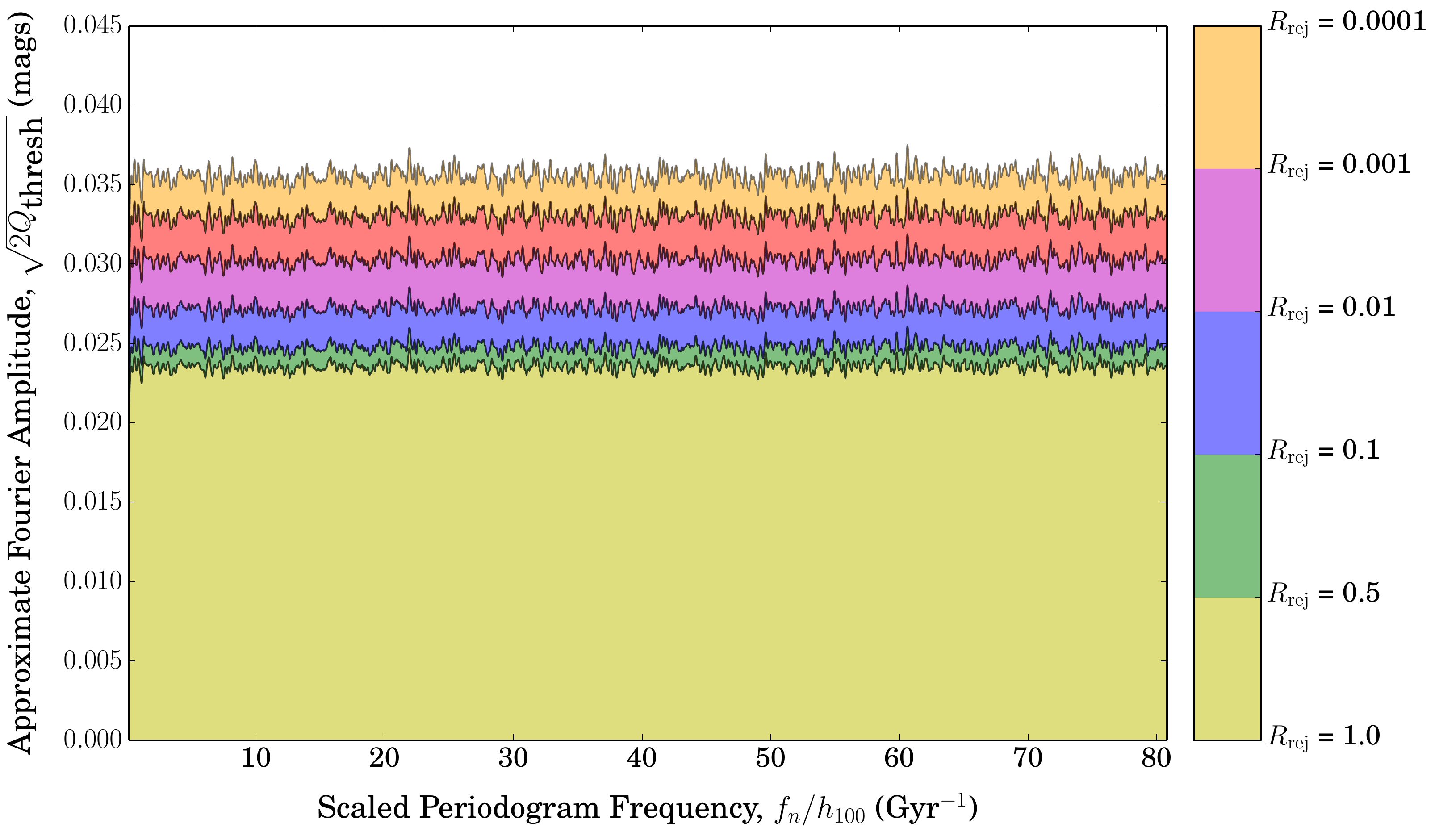}
\caption{\label{fig:ratioFourierLimits}  The Fourier amplitude limits for a given probability ratio measured with respect to the most excursive (i.e. least likely) Fourier mode of the Pantheon SNe Ia.  The contours indicate the rejection regions for various values of $R_{\textrm{rej}}$.  We applied a rectangular smoothing of width $10  f_{\textrm{min}}$ to the threshold Fourier amplitudes.  By comparing the Fourier spectrum of the distance modulus residuals of a particular ACM to these rejection regions, we determine how extreme the considered ACM is relative to the Pantheon SNe Ia.  Any ACM that predicts distance modulus residuals with a Fourier amplitude at or above the $R_{\textrm{rej}} = 0.0001$ contour (the transition from orange to white, about 36 millimags at most frequencies) is $\geq 10000$ times less likely to arise by random chance from the $\Lambda$CDM cosmology than the most extreme mode of the Pantheon SNe Ia.  Such ACMs are thus strongly ruled out by Fourier analysis.  The amplitude constraints are slightly stronger at the lowest 0.1\% of considered frequencies, with the $R_{\textrm{rej}} = 0.0001$ constraint at about 31 millimags when $f_n = f_{\textrm{min}}  \simeq 0.08 \ \textrm{Gyr} ^ {-1} h_{100}$. }  
\end{figure*} 

The first statistic that we consider is the reduced $\chi^2$ statistic, $r \chi ^2_{\nu}$, defined as 
\begin{equation}
    \begin{aligned} 
        r \chi^2_{\nu} = \frac{\chi^2}{\nu} 
                               = \sum \limits_{i } \frac{ (\Delta \mu_{i,\textrm{theoretical}} - \Delta \mu_{i,\textrm{observed}}) ^ 2} {\nu \sigma_i ^ 2} \ ,
     \end{aligned} 
\end{equation}
where $\chi^2$ is the standard $\chi^2$ statistic, $\nu \equiv $ \textit{(\# of data points)} $-$ \textit{(\# of model free parameters)} is the total number of degrees of freedom, and the sum is taken over all observed SNe.  Our data set consists of $1048$ data points.  Each considered ACM has a number of free parameters equal to the number of constrainable parameters of the model plus the overall mean of the predicted $\Delta \mu$, which is subtracted from each data set.  For Equations \ref{eq:wOfTDef}, \ref{eq:rhoDEAOfTDef}, \ref{eq:rhoDEBOfTDef}, and \ref{eq:GOfTDef}, $\nu$ is equal to, respectively, $1044$, $1044$, $1043$, and $1042$. 

The likelihood of randomly drawing an $r \chi^2_{\nu}$ value from a true $r\chi^2_{\nu}$ distribution that is larger than the $r \chi^2_{\nu}$ value associated with some ACM is given by 
\begin{equation} 
    \begin{aligned} 
    P(r \chi^2_{\nu} & \geq r \chi^2_{\nu, \textrm{ACM}} ) = \int_{ r \chi^2_{\nu, \textrm{ACM}} \nu }^{\infty} d (\chi^2) p_{\nu}(\chi^2)  \ ,
    \end{aligned} 
\end{equation}
where $p_{\nu}(\chi^2) = 1/(2 ^{\nu / 2} \Gamma(\nu/2)) e ^{-\chi^2 /2} (\chi^2)^{\nu/2 -1}$ is the probability density of the $\chi^2$ distribution, and where $r \chi^2_{\nu, \textrm{ACM}}$ is the $r \chi^2_{\nu}$ value of the ACM.  
The probability ratio of randomly acquiring $r \chi^2_{\nu, \textrm{ACM}}$ vs randomly acquiring the $r \chi^2_{\nu}$ value of the canonical universe is 
\begin{equation} \label{eq:chiSqrRejDef}
R_{r \chi^2_{\nu}} = \frac{ \int_{r \chi^2_{\nu, \textrm{ACM}} \nu}^{\infty} d (\chi^2) p_{\nu}(\chi^2) }{ \int_{r \chi^2_{\nu, \textrm{can}} \nu}^{\infty} d (\chi^2) p_{\nu}(\chi^2) }  \ ,
\end{equation}
where $r \chi^2_{\nu, \textrm{can}}$ is the $r \chi^2_{\nu}$ of the canonical universe: 
\begin{equation}
r \chi^2_{\nu, \textrm{can}} = \sum \limits_{i} \frac{ (0 - \Delta \mu_{i,\textrm{observed}}) ^ 2} {\nu \sigma_i ^ 2} \ .
\end{equation}

For a specified rejection ratio, $R_{\textrm{rej}}$, we define the threshold $r \chi^2_{\nu}$ value, $r \chi^2_{\nu,\textrm{thresh}}$, by  
\begin{equation} \label{eq:chiSqrRejDef}
    \begin{aligned} 
     &\int_{r \chi^2_{\nu, \textrm{thresh}} \nu}^{\infty} d (\chi^2) p_\nu(\chi^2) \\
     &= \Big ( \int_{r \chi^2_{\nu, \textrm{can}} \nu}^{\infty} d (\chi^2) p_\nu(\chi^2) \Big ) \ R_{\textrm{rej}} \\ 
     & \simeq 0.467 \ R_{\textrm{rej}}   \ .
     \end{aligned} 
\end{equation}
Any ACM with a $r \chi^2_{\nu}$ value larger than $r \chi^2_{\nu, \textrm{thresh}}$ is deemed inconsistent with the data for the choice of $R_{\textrm{rej}} $.

The second statistic that we consider is based on the Fourier analysis of Section \ref{sec:statMethodFourier}.  
For a particular ACM with periodogram values $Q_{n, \textrm{ACM}}$, the probability that one or more periodogram values measured in the canonical universe would be more extreme than the most extreme value of $Q_{n, \textrm{ACM}}$ is 
\begin{equation}
    \begin{aligned} 
P( \textrm{a single} & \textrm{ realization of $\Lambda$CDM} \\
    & \ \ \ \textrm{ more extreme than ACM}) \\
    & = 1 - (\textrm{max}(r_n(Q_{n,\textrm{ACM}}))) ^{N_{\textrm{peak}}}  \ .
    \end{aligned} 
\end{equation}  
The probability ratio of randomly acquiring the most extreme periodogram value associated with an ACM vs randomly acquiring the most extreme periodogram value associated with the Pantheon data set is 
\begin{equation} 
    \begin{aligned} 
R &_{\textrm{Fourier}} =\frac{ 1 - (\textrm{max}(r_n(Q_{n,\textrm{ACM}}))) ^{N_{\textrm{peak}}} }{ 1 - (\textrm{max}(r_m(Q_{m,\textrm{Pantheon}}))) ^{N_{\textrm{peak}}} } \ , 
    \end{aligned} 
\end{equation}
where the values of $Q_{n,\textrm{Pantheon}}$ are shown in Figure \ref{fig:trueFourierDecomp}.  
For a specified rejection ratio, we define the threshold periodogram value at frequency $f_n$, $Q_{n,\textrm{thresh}}$, by 
\begin{equation} \label{eq:fourierRejDef}
    \begin{aligned} 
1 &- (r_n(Q_{n,\textrm{thresh}})) ^{N_{\textrm{peak}}} \\
   & = (1 - (\textrm{max}(r_m(Q_{m,\textrm{Pantheon}}))) ^{N_{\textrm{peak}}} ) R_{\textrm{rej}}  \\ 
   & \simeq  \ 0.284 \ R_{\textrm{rej}} \ .
    \end{aligned} 
\end{equation}
If any periodogram value of a given ACM exceeds the corresponding value of $Q_{n, \textrm{thresh}}$, then we deem that ACM inconsistent with the data for that choice of $R_{\textrm{rej}}$.  We show the values of $Q_{n, \textrm{thresh}}$ for some choices of $R_{\textrm{rej}}$ in Figure \ref{fig:ratioFourierLimits}.  

Typically, Equations \ref{eq:chiSqrRejDef} and \ref{eq:fourierRejDef} are solved numerically. 

We emphasize the generality of both of these methods.  Just as the reduced $\chi^2$ statistic can be used to test the consistency of any ACM with observed residuals, so too can the Fourier spectrum of the residuals of any ACM be computed and compared to the Fourier limits shown in Figure \ref{fig:ratioFourierLimits}.  The constraints that we discuss in Section \ref{sec:resultsConstrainingModels} on the models described in Section \ref{sec:altModels} provide illustrative examples of using these techniques and should not be regarded as exhaustive. 

\section{RESULTS} \label{sec:res}

\subsection{Assessing the Consistency of the Pantheon SNe with $\Lambda \mathrm{CDM}$ } \label{sec:resultsFourier}
\begin{figure*}
\centering
\includegraphics[width=0.7\textwidth]{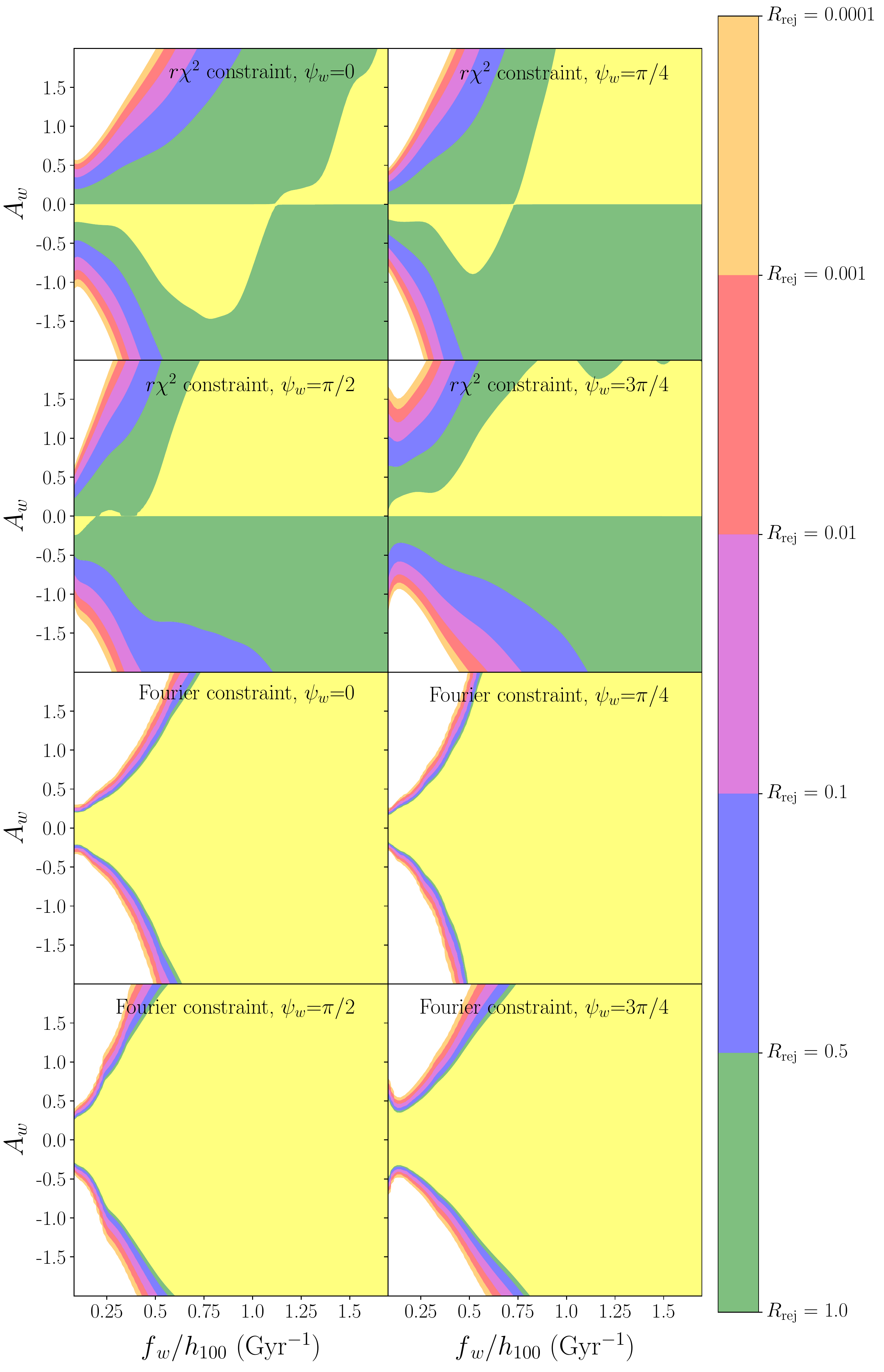} 
\caption{\label{fig:fromWRChiSqr}  The portions of the parameter space of the ACM characterized by the DE EoS Parameter of Equation \ref{eq:wOfTDef} that are rejected by the reduced $\chi^2$ technique (top 2 rows) and the Fourier constraining technique (bottom 2 rows) for various rejection probability ratios, $R_{\textrm{rej}}$.  As discussed in Section \ref{sec:constrainingMethods}, a model is rejected by a given statistical test if it is $\geq 1/R_{\textrm{rej}}$ times less likely to occur than the null hypothesis of that test.  A smaller value of $R_{\textrm{rej}}$ corresponds to the rejection of a smaller portion of a model's parameter space.  The distance modulus residuals of this ACM are characterized by oscillations that decay rapidly with $\tau$, with the characteristic size of the oscillations diminishing quickly with increasing $f_w$ (see Figure \ref{fig:muResidsFromWofT}).  Although the Fourier spectra of these residuals do contain a peak at $f_w$, that peak is typically dwarfed by the low order Fourier structure characterizing the decay of the $\Delta \mu$ oscillations.  The Fourier constraints above derive from constraining this low order structure, and are typically similar to the constraints of the standard reduced $\chi^2$ analysis.  Because the distance modulus residuals predicted by the ACM of Equation \ref{eq:wOfTDef} oscillate only minutely in $\tau$ relative to their other structure, Fourier analysis does not yield a stronger constraint than standard statistical tests.   }
\end{figure*}

We now discuss the results of using the Fourier analysis of Section \ref{sec:statMethodFourier} to check the consistency of the Pantheon data set with the predictions of the canonical Planck cosmology.  

As discussed in Section \ref{sec:statMethodFourier}, we consider $N_f = 4995$ evenly spaced frequencies between $f_{\textrm{min}}  \simeq 0.08 \ \textrm{Gyr} ^ {-1} h_{100}$ and $f_{\textrm{max}} \simeq 80.8 \  \textrm{Gyr} ^ {-1} h_{100}$.  At each frequency, we approximate the Fourier power by computing the periodogram defined in Equation \ref{eq:periodogramDef} for our observed values of $\Delta \mu$ and $\tau$.  We simulate $N_r = 1000$ random data sets from an assumed null relation between $\Delta \mu$ and $\tau$, subject to the reported errors.  For each frequency, we use Equation \ref{eq:maxperiodogramDef} to determine the maximum permissible periodogram value given a specified $P_{\textrm{rej}}$.  We then determine the approximate Fourier amplitude from the periodogram values using Equation \ref{eq:fourierPowerToAmp}

In Figure \ref{fig:trueFourierDecomp}, we plot the measured Fourier amplitudes and the threshold Fourier amplitudes for various choices of $P_{\textrm{rej}}$.  
The measured peak that is most excursive from the canonical cosmology is located at frequency $f_{4349}  \simeq 70.37 \ \textrm{Gyr} ^ {-1} h_{100} $ and has an approximate Fourier amplitude of about 
\begin{equation}
\sqrt{2 Q_{4349,{\textrm{Pantheon}}}} \simeq \sqrt{2  \ 0.000291} \simeq 0.024 \ \textrm{mags} \ .
\end{equation}  
There is a 
\begin{equation}
1-r_{4349}(Q_{4349,{\textrm{Pantheon}}} = 0.000291) \simeq 0.000669
\end{equation} chance that the value of $Q_{4349,{\textrm{Pantheon}}}$ would be larger than it is found to be if the Pantheon data set resulted from the Planck $\Lambda$CDM cosmology.  With $N_{\textrm{peak}} = 500$ presumably independent peaks, there is a 
\begin{equation}
1-(1 - 0.000669)^{500} \simeq 0.284
\end{equation}
chance that a single realization of the canonical cosmology would have at least one peak that is more extreme than the most extreme peak characterizing the Pantheon data.   

Those are not long odds, and so we conclude that our observed periodogram is consistent with the distribution of periodograms that could be measured from a null relation between $\Delta \mu$ and $\tau$.  The Pantheon data set remains consistent with the predictions of $\Lambda$CDM when subject to Fourier analysis.  

\subsection{Constraints on Alternate Cosmological Models} \label{sec:resultsConstrainingModels}

We now constrain the three specific ACMs discussed in Section \ref{sec:altModels} using the reduced $\chi^2$ and Fourier statistical analyses of Section \ref{sec:constrainingMethods}.   

For the ACM characterized by the oscillatory DE EoS parameter of Equation \ref{eq:wOfTDef}, the low order structure that dominates this model's predicted relations between $\Delta \mu$ and $\tau$ (see Figure \ref{fig:muResidsFromWofT}) broadens and flattens the resulting Fourier spectrum.  This broadening significantly weakens the constraining power of the Fourier method, with the strongest such constraints resulting from the lowest searched for frequencies.  The Fourier constraining method and the $r \chi^2_{\nu}$ methods offer similar constraints on this model's allowable parameter ranges, and we show the constraints furnished by both methods in Figure \ref{fig:fromWRChiSqr}. 
 
 For frequencies, $f_w$, around $0.1 \ \textrm{Gyr} ^ {-1} h_{100}$, we find that oscillation amplitudes, $A_{w}$, larger than around $0.5$ are strongly ruled out.  As $f_w$ increases, the constraints on $A_w$ rapidly weaken due to the double-integral relation between $d_L$ and $w_{\Lambda}$.  Oscillations in the value of $w_{\Lambda}$ have relatively small effects on the evolution of $d_L$, with the effect growing smaller as the oscillations become more rapid.  The power of SNe Ia observations to constrain oscillations in $w_{\Lambda}$ weakens as the hypothetical frequency of oscillation grows. 

\begin{figure*}
\centering
\includegraphics[width=0.7\textwidth]{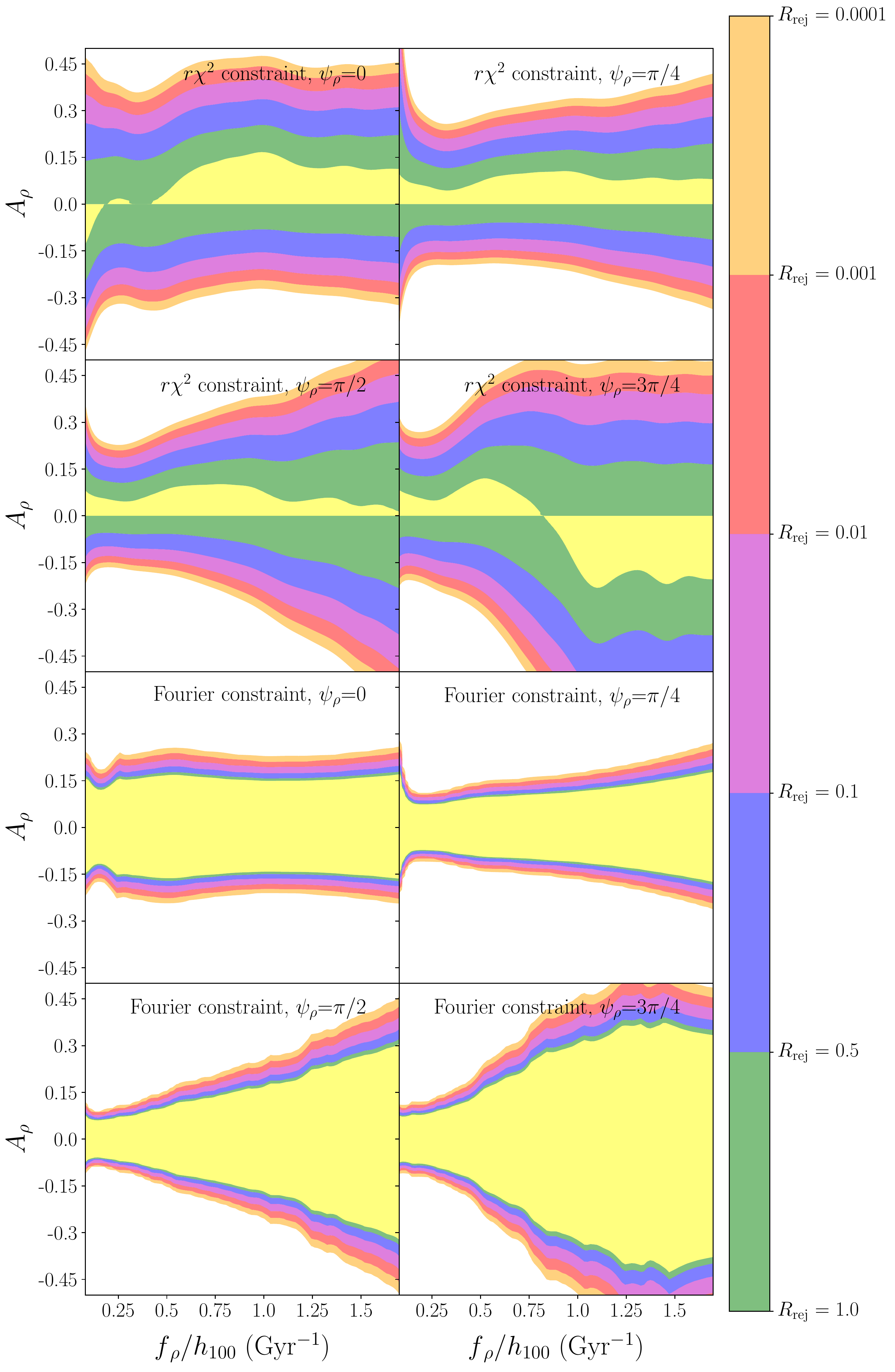}
\caption{\label{fig:fromDEAFourierLims} As in Figure \ref{fig:fromWRChiSqr} except that the ACM considered is that characterized by the oscillatory DE energy density of Equation \ref{eq:rhoDEAOfTDef}.  The distance modulus residuals of this ACM are characterized by decaying oscillations, and the most recent SNe produce the largest expected deviation from the predictions of $\Lambda$CDM (see Figure \ref{fig:muResidsFromDEAofT}).  The Fourier spectra of these residuals are generally characterized by two peaks: one at the lowest measured frequencies and one at $f_\rho$.  The relative sizes of these peaks change with both $\psi_\rho$ and $f_\rho$, and the Fourier constraint for a given set of parameters is based on the larger of these two peaks.  For example, the low frequency peak is generally larger than the $f_\rho$ peak for $\psi_\rho = \pi / 4$, meaning that the Fourier constraint for this phase grows relatively slowly as $f_\rho$ increases.  In contrast, the low frequency peak is very small when $\psi_\rho = 3 \pi / 4$ and the Fourier constraint for this phase weakens as $f_\rho$ increases because the $f_\rho$ peak diminishes as $f_\rho$ increases.  } 
\end{figure*}

\begin{figure*}
\centering
\includegraphics[width=0.7\textwidth]{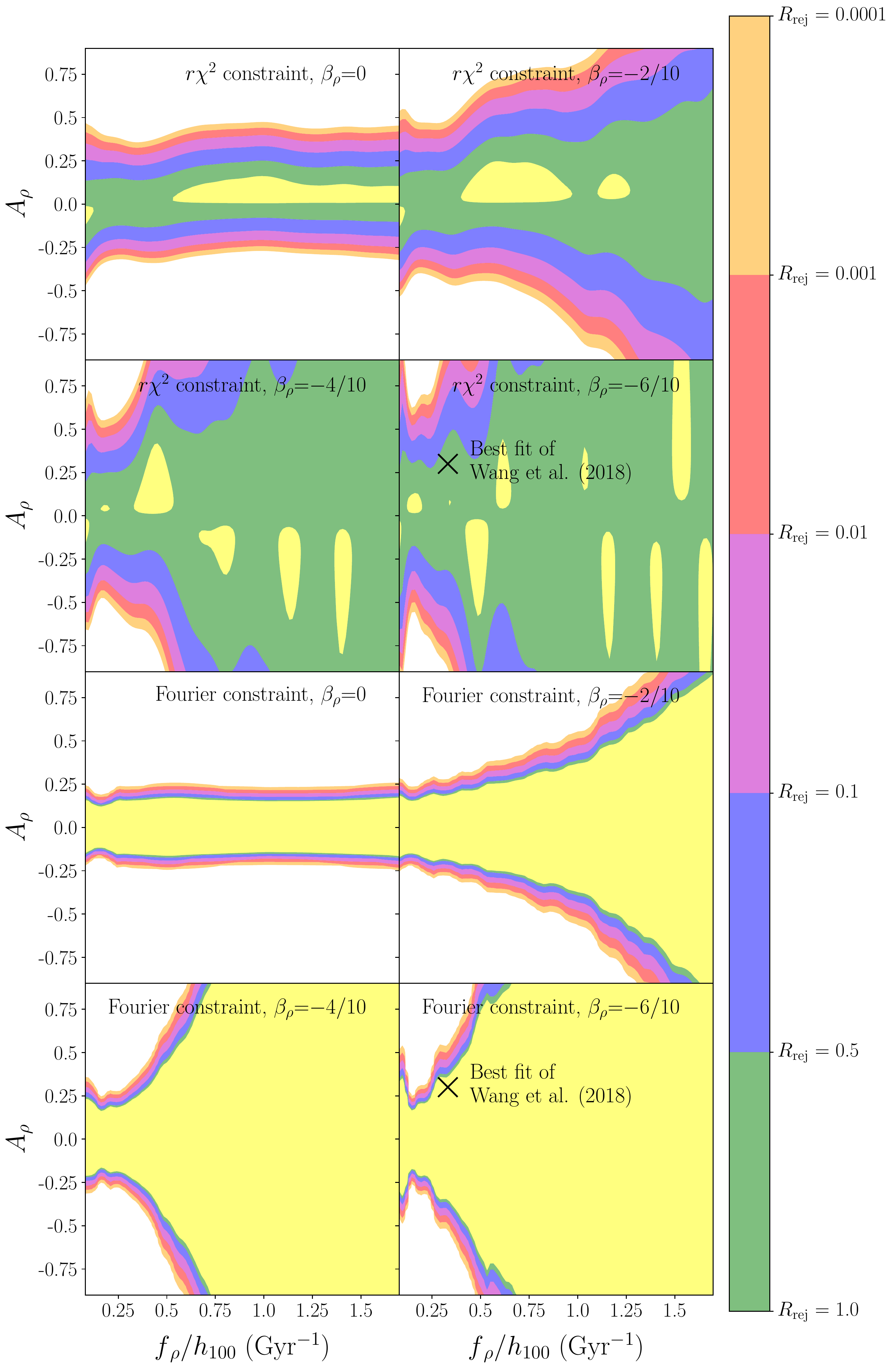} 
\caption{\label{fig:fromDEBFourierLims}  As in Figure \ref{fig:fromWRChiSqr} except that the ACM considered is that characterized by the decaying oscillatory DE energy density of Equation \ref{eq:rhoDEBOfTDef}.  We have fixed $\psi_\rho = 0$ and $\tau_\rho \simeq 6.2/ h_{100} \textrm{Gyr} $. 
We find that the following parameter choice produces an $X(z)$ function that appears similar, though not identical, to the `all-data' $X(z)$ function shown in panel (F) of Figure 1 of \cite{Wang18}: $(A_\rho, f_\rho, \beta_\rho, \psi_\rho, \tau_\rho) = (0.3, 0.034 \mathrm{Gyr}^{-1},-0.6, 0.0, 9.2\mathrm{Gyr})$.  This point, marked with a cross above, is consistent with the data, according to both the reduced $\chi^2$ and Fourier analyses.}  
\end{figure*} 

\begin{figure*}
\centering
\includegraphics[width=0.7\textwidth]{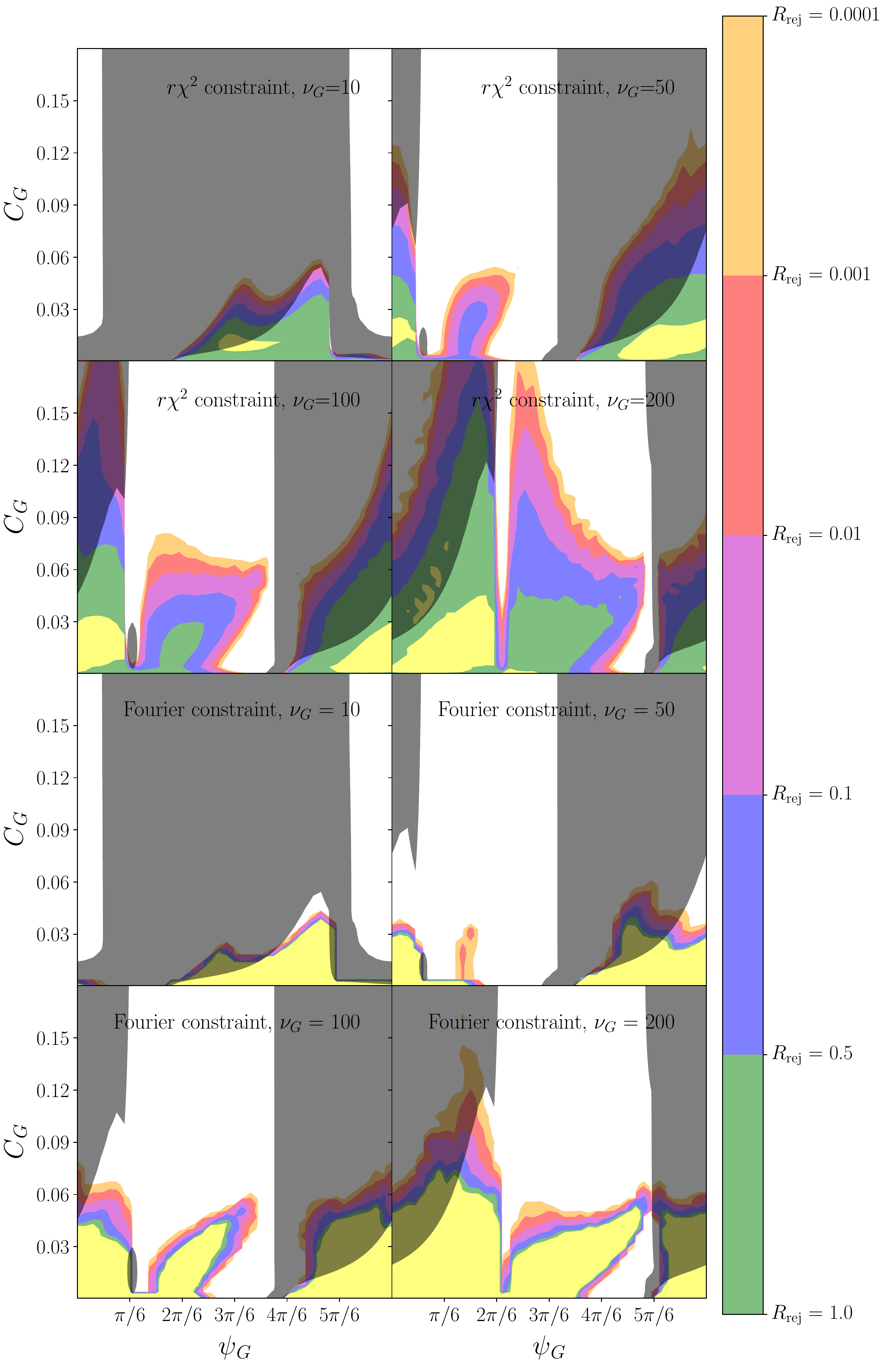}
\caption{\label{fig:fromGFourierLims}  As in Figure \ref{fig:fromWRChiSqr} except that the ACM considered is characterized by a scalar field of gravity evolving according to the oscillatory coupling of Equation \ref{eq:GOfTDef}.  Here, $A_G = 7$, $\epsilon_G$ has been tuned so as to minimize the contemporary value of $dG/dt$, and $\phi_0$ has been tuned so that $G(\phi_0) = G_0$.  We recalculate the tuning of $\epsilon_G$ and $\phi_0$ for every considered set of constrainable parameter values.
 The $5 \sigma$ limits on $(dG/dt)/G$ and on $(d^2G/dt^2)/G$ from the lunar ranging experiment (see Appendix \ref{app:lunarGConstraints}), are shown as dark shading. 
Reduced $\chi^2$ analysis, Fourier analysis, and contemporary measurements of the derivates of $G$ offer complementary constraints.  The portion of the model parameter space that predicts a large contemporary value of $dG/dt$ is more effectively constrained by direct measurements of $G$ and its derivatives.  For the portion of the parameter space that predicts a currently stationary value of $G$, reduced $\chi^2$ and Fourier analysis of SNe Ia provide stronger constraints than contemporary measurements of $G$.  } 
\end{figure*}

As is visible in Figure \ref{fig:muResidsFromDEAofT}, the distance modulus residuals resulting from the simply oscillating DE energy density of Equation \ref{eq:rhoDEAOfTDef} are characterized by both a decaying envelope and highly coherent oscillations.  The Fourier spectra of these residuals is characterized by two significant structures: the spectrum of the decaying envelope and the isolated Fourier peak at the oscillation frequency, $f_\rho$.  For small values of $f_\rho$, the large size of the $\Delta \mu$ oscillations leads to a large Fourier peak at $f_\rho$ , creating a strong Fourier constraint.  However, as the frequency increases, the size of the $\Delta \mu$ oscillations diminish.  For higher values of $f_\rho$, the strongest Fourier constraints derive from the Fourier structure of the decaying envelope.  The Fourier constraining method provides a stronger constraint than the reduced $\chi^2$ method, at all considered values of $f_\rho$.  We show both sets of constraints in Figure \ref{fig:fromDEAFourierLims}.  

We consider Equation \ref{eq:rhoDEBOfTDef} to replicate the slowly oscillating DE energy density that \cite{Wang18} claim is a better match to the the available cosmological data than the standard $\Lambda$CDM cosmology.  We fix $\psi_\rho = 0$ and $\tau_\rho$ to the middle of the canonical $\Lambda$CDM $\tau$ window spanned by the Pantheon SNe Ia: $\tau_\rho \simeq 6.2/ h_{100} \textrm{Gyr} $.  When $\beta_\rho < 0$, the amplitude of the $\Delta \mu$ oscillations predicted by this ACM decay more slowly then those of the simply oscillating DE energy density of Equation \ref{eq:rhoDEAOfTDef}, and the frequency of the oscillations also decreases in time.  The slower decay time reduces the low frequency Fourier structure associated with the decay envelope, and the changing oscillation frequency broadens and shrinks the Fourier peak of the oscillations.  Fourier analysis places weaker constraints on the amplitude of the ACM described by Equation \ref{eq:rhoDEBOfTDef} relative to that of Equation \ref{eq:rhoDEAOfTDef}.    

We find that, when $A_\rho = 0.3$, $f_\rho = 0.034\ \textrm{Gyr}^{-1} h_{100}$, and $\beta_\rho = -0.6$ (marked with an `X' in Figure   \ref{fig:fromDEBFourierLims}), Equation \ref{eq:rhoDEBOfTDef} resembles the $X(z)$ function shown in panel F of Figure 1 of \cite{Wang18}.  Both reduced $\chi^2$ and Fourier analysis find that Equation \ref{eq:rhoDEBOfTDef} is consistent with the data for this choice of parameters.  
However, these parameter values are near the edge of parameter space that Fourier analysis deems consistent with the data.  This cosmology produces a peak in the Fourier spectrum of the SNe Ia distance modulus residuals that is almost as excursive as the most excursive Fourier peak in Figure \ref{fig:trueFourierDecomp}.  A modest increase in the size of the used SNe Ia data set would be sufficient to either confirm or confidently rule out the oscillating DE energy density claimed by \cite{Wang18}.

The variations in $G$ that arise from the evolving scalar field of Equation \ref{eq:GOfTDef} are highly sensitive to our choice of parameters.  And because $d_L$ depends on $G$ through both an integral relation and through a direct power law relation (see Appendix \ref{app:fromGODEs}) these variations manifest strongly in the predicted values of $\Delta \mu$.  Because the predictions of $\Delta \mu$ are highly dependent on our choice of phase (see Figure \ref{fig:muResidsFromGofT}), we densely sample $C_G$ and $\psi_G$ over a small set of fixed $\nu_G$ values. 

Those parameter choices that produce $\Delta \mu$ values with a significant polynomial component (e.g. $C_G = 0.04$, $\psi_G=2\pi/3$, $\nu_G = 100$) are effectively ruled out with reduced $\chi^2$ analysis.  Parameter choices with $\Delta \mu$ oscillations that are large relative to their polynomial drift (e.g. $C_G = 0.04$, $\psi_G=0$, $\nu_G = 100$) are more strongly constrained by Fourier analysis.  We show both sets of constraints, along with the Lunar Ranging constraints of Equation \ref{eq:myLunarConstraints}, in Figure \ref{fig:fromGFourierLims}. 

As discussed in Appendix \ref{app:fromGODEs}, the SNe Ia residuals predicted by a model with a varying $G$ are dependent on the strength of the dependence of $L$ on $G$.  If the true dependence is more or less extreme than the $L \propto G ^ {-3/2}$ relation proposed by \cite{Gaztanaga01}, a possibility considered by \citet{Wright18}, than our measured constraints will be stronger or weaker, respectively.  

Because its behavior is so sensitive to our choice of model parameters, the ACM described by Equation \ref{eq:GOfTDef} illustrates the power of using complimentary techniques to constrain the parameter space of a single model.  The combined application of reduced $\chi^2$ analysis, Fourier analysis, and contemporary measurements of $G$ and its derivatives rule out much of the considered parameter space even though there are sections of the parameter space where each method is individually weak.   

\section{DISCUSSION AND CONCLUSION} \label{sec:discussion}

We have used the Pantheon data set of SNe Ia to search for deviations from the predictions of $\Lambda$CDM that oscillate in conformal time.  We have taken advantage of the size of the Pantheon data set to search for deviations that occur over redshift ranges as small as about $0.05$ and as large as the full observed redshift range of about $2.3$.  To maximize the power of our search, we developed a frequency based analysis method that should be regarded as supplementary to the standard $\chi^2$ analysis.  

We first applied our analysis in a model-agnostic way and determined that there is roughly $28.4\%$ chance that data drawn from the Planck $\Lambda$CDM cosmology would exhibit at least one Fourier peak that is more extreme than that most extreme Fourier peak belonging to the Pantheon data set.  We thus concluded that the observed Fourier spectrum is consistent with the distribution of Fourier spectra expected from the canonical $\Lambda$CDM cosmology.  The Pantheon SNe Ia data set exhibits no evidence of deviations from $\Lambda$CDM that oscillate in conformal time.  

We then discussed how both the reduced $\chi^2$ statistic and Fourier analysis can be used to constrain proposed alternate cosmological models (ACMs).  
To illustrate the complementary power of these two statistical tests, we considered and constrained several candidate ACMs: one in which the DE EoS parameter oscillates in conformal time around $-1$, two versions of an ACM in which the DE energy density oscillates around its canonical value, and one in which gravity arises from a scalar field that evolves under an oscillatory coupling.

 We determined that the strongest constraints on the amplitude of the oscillatory DE EoS parameter described by Equation \ref{eq:wOfTDef} are generally obtained at low frequencies, with the strongest constraint being a roughly 99.9\% certainty rejection of amplitudes, $A_w$  larger than about $0.5$ for phases, $\psi_w$ of either $0$ or $\pi/4$ with frequencies, $f_w$, between $0.08 \  \textrm{Gyr} ^ {-1} h_{100}$ and $0.2 \ \textrm{Gyr} ^ {-1} h_{100}$.  
 The $f_w$ location of the strongest $A_w$ constraint is somewhat $\psi_w$ dependent, but the constraints on $A_w$ do rapidly weaken in $f_w$ for all $\psi_w$ when $f_w \geq 0.3 \ \textrm{Gyr} ^ {-1} h_{100}$.  
 The rapid weakening of this constraint is a consequence of the double integral relation between $d_L$ and $w_\Lambda$.  
 
We found that Fourier analysis provides the strongest constraints on the amplitude, $A_\rho$, of the oscillatory DE energy density described by Equation \ref{eq:rhoDEAOfTDef}.  The strongest constraints typically rule out values of $A_\rho$ larger than about $0.2$ for values of $f_\rho$ between $0.08 \ \textrm{Gyr} ^ {-1} h_{100}$ and $0.3 \ \textrm{Gyr} ^ {-1} h_{100}$, though the precise frequency ranges and amplitude constraints are $\psi_\rho$ dependent.  The speed with which the $A_\rho$ constraint weakens as $f_\rho$ grows is also $\psi_\rho$ dependent due to the relation between $\psi_\rho$ and the relative strength of the primary features of the Fourier spectrum. 

We considered the decaying, oscillatory DE energy density described by Equation \ref{eq:rhoDEBOfTDef} primarily to test the claim of \cite{Wang18} that a slightly oscillating cosmology is a better match to the available cosmological data than the canonical $\Lambda$CDM cosmology.  
For nonzero values of the decay parameter, $\beta_\rho$, this ACM's frequency of oscillation changes in conformal time and the associated Fourier peak broadens and diminishes.  We find that, for $\beta_\rho$ values less than $0$, the constraining power of both constraining methods decreases as the frequency, $f_\rho$, increases.  The rate of the decrease depends on the value of $\beta_\rho$. 
The strongest $A_\rho$ constraint of $0.25$ is found around $f_\rho = 0.25 \textrm{Gyr} ^ {-1} h_{100}$ for all $\beta_\rho$. 

By applying both the reduced $\chi^2$ and Fourier constraining methods, we found that our best approximation of the oscillating cosmology shown in \cite{Wang18} is consistent with the Pantheon SNe Ia.  However, this approximation is near the boundary of the portion of the parameter space that Fourier analysis rules out.  Thus, small corrections to the fit parameters or a modest increase in the size of the SNe Ia data set could lead to the exclusion or confirmation of this ACM.

Because of the direct power law dependence of the SNe Ia luminosity on the local gravitational constant, an ACM in which gravity derives from an evolving scalar field produces significant excursions in the distance moduli of SNe Ia.  For the particular ACM described by Equation \ref{eq:GOfTDef} in which the scalar coupling oscillates, we found that some small subset of model parameters predict a currently stationary value of $G$.  
In this narrow parameter regime, both reduced $\chi^2$ and Fourier analyses of distance SNe Ia provide stronger constraints than contemporary measurements of $G$ and its derivatives.  

Improperly diagnosed systematic errors may plague the Pantheon data set \citep{Scolnic18} and could produce spurious Fourier signals.  As the Fourier analysis technique introduced here is further refined, the possible impact of such systematic errors can and should be properly accounted for.  Future efforts could integrate these unaccounted for systematics into the determination of the distribution of Fourier modes that could result from the $\Lambda$CDM cosmology by including these unaccounted for systematics in the generation of the artificial data.  The consistency of the Fourier profile of the observed data with $\Lambda$CDM and the Fourier constraints placed on a given ACM could then be determined following the same techniques outlined in this work.  

Frequency based analysis is particularly apt at constraining those ACMs that predict coherent, oscillatory deviations from the predictions of $\Lambda$CDM.  
Future proposed ACMs should conform to the constraints of both the standard $\chi^2$ test and of Fourier analysis. 
For any ACM that predicts an oscillatory component in the expansion history of the universe, the Fourier spectrum of the residuals of the ACM should be computed and compared to the limits of Figure \ref{fig:ratioFourierLimits}.  The most extreme mode of the Fourier spectrum determines, for a particular choice of $R_{\textrm{rej}}$, if a model is consistent with the available observations.  
In particular, those ACMs that predict distance modulus residuals with a Fourier amplitude larger than about 36 millimags at any frequency larger than $0.08\ \textrm{Gyr}^ {-1} h_{100}$ are ruled out with about $99.99\%$ confidence.  

Some ACMs predict oscillatory deviations from $\Lambda$CDM with amplitudes that are too small to be constrained by Fourier analysis given the currently available data.  For example, we attempted to place meaningful constraints on the `monodromic dark energy' proposed by \cite{Schmidt17}.  Although that oscillating dark energy model does produce a peak in the Fourier spectrum of the predicted SNe Ia distance modulus residuals, the height of the peak is about a factor of $5$ smaller than the threshold Fourier amplitudes of Figure \ref{fig:ratioFourierLimits}. 

The addition of future SNe Ia observations will improve the constraints derived in this work, enabling the confirmation or rejection of models that are presently consistent with the data. 
For the ideal case of SNe Ia with even $\tau$ spacing, the maximum allowable periodogram value, $Q_{\textrm{max},n}$, for a given rejection probability, $P_{\textrm{rej}}$, roughly scales as $Q_{\textrm{max},n} \propto (\sigma_{\textrm{SN}} ^2)/( N_{\textrm{SN}})$, 
where $\sigma_{\textrm{SN}}$ is the typical SNe Ia distance modulus uncertainty and $N_{\textrm{SN}}$ is the total number of SNe.  We thus expect the constraints on maximum allowable Fourier amplitudes to scale as $\propto \sqrt{(\sigma_{\textrm{SN}} ^2)/ N_{\textrm{SN}}}$.  During the first years of its operation, the Large Synoptic Survey Telescope (LSST) is expected to observe about $300$ times as many SNe Ia as currently comprise the Pantheon data set, with comparable uncertainties.  If such a vision is realized, than the LSST should move the 99.99\% constraint of Figure \ref{fig:trueFourierDecomp} to roughly $\sqrt{1^2/300} \ 36  \textrm{ millimags}\simeq 2 \ \textrm{millimags}$, enabling astronomers to either detect or rule out any cosmological signals with Fourier components larger than this threshold.  

\vspace{5mm}
We are grateful to the US Department of Energy for their support under award DE-SC0007881. 

\appendix
\renewcommand{\theequation}{A\arabic{equation}} 

\section{ACMS WITH AN EVOLVING DARK ENERGY EQUATION OF STATE} \label{app:fromWODEs}
Here, we elaborate on the derivation of the ODEs corresponding to an ACM characterized by an evolving DE EoS parameter, $w_{\Lambda} \rightarrow w_{\Lambda}(\tau, z)$. 
Our goal in this Appendix is to derive a set of coupled ODEs that can be numerically solved to describe the evolution of cosmic parameters for a given expression for $w(\tau,z)$.  

We begin by deriving some useful differential equations that follow immediately from definitions: 
\begin{equation}\label{eq:tOfZODE}
    \begin{aligned} 
    \frac{dt}{dz}  = \frac{dt}{da}\frac{da}{dz} = \frac{1}{da/dt} \frac{d(1/(1+z))}{dz} 
                       =  \frac{a(1+z)}{da/dt} \frac{-1}{(1+z)^2} = - \frac{1}{H}\frac{1}{1+z} \ ,
    \end{aligned} 
\end{equation}
\begin{equation}\label{eq:tauOfZODE}
\tau = \int_0^z dz' \frac{1}{H(z')} \Rightarrow \frac{d \tau}{dz} = \frac{1}{H(z)} \ ,
\end{equation}
and 
\begin{equation}\label{eq:dLOfZODE}
    \begin{aligned} 
 d_L  = c(1+z) \int_0^z dz' \frac{1}{H(z')} 
 \Rightarrow \frac{d\ d_L}{dz}= c \int_0^z dz' \frac{1}{H(z')} + c \frac{c(1+z)}{H(z)} 
 =  \frac{d_L}{(1+z)} +  \frac{c(1+z)}{H(z)} \ ,
    \end{aligned} 
\end{equation}
where, in our expression for $d(d_L) /dz$, we have implicitly assumed that SNe Ia behave as expected.  

We now move to the Friedmann Equation for a flat universe: 
\begin{equation} \label{eq:friedmann}
    \begin{aligned} 
 H^2 = \frac{8 \pi G }{3 c ^2} (\rho_M + \rho_{\Lambda} + \rho_R)  
 \Leftrightarrow \frac{H^2}{H_0^2} = (\Omega_m + \Omega_{\Lambda} + \Omega_r) \ ,
    \end{aligned} 
\end{equation}
where $\rho_n$ is the energy density of universe constituent $n$, and $\Omega_n$ is $\rho_n$ scaled by the present day critical energy density, $\rho_{c,0} = (3H_0^2 c ^ 2)/(8 \pi G)$.  

The fluid equation provides an ODE for each $\rho_n$: 
\begin{equation} \label{eq:fluid} 
    \begin{aligned}
\frac{d\rho_n}{dt}  = -3 H (\rho_n + P_n ) 
                             = -3 \frac{da/dt}{a} \rho_n(1+ w_n ) \ ,
    \end{aligned} 
\end{equation}
where we have used the definition of the EoS parameter, $w_n$, to express $P_n$ in terms of $\rho_n$.  

Using Equations \ref{eq:friedmann} and \ref{eq:fluid}, we can determine the evolution of matter, for which $w_M = 0$:
\begin{equation} \label{eq:rhoMOfZ}
    \begin{aligned} 
 \frac{d\rho_M}{dt} = -3 \frac{da/dt}{a} \rho_M 
 \Rightarrow \rho_M = \frac{1}{a^3} \rho_{M,0} = (1+z)^3\rho_{M,0} \ ,
    \end{aligned} 
\end{equation}
and of radiation, for which $w_R = 1/3$:
\begin{equation} \label{eq:rhoROfZ}
    \begin{aligned} 
\frac{d\rho_R}{dt} = -3 \frac{da/dt}{a} \rho_R(1 + \frac{1}{3}) 
 \Rightarrow \rho_R = \frac{1}{a^4} \rho_{r,0}= (1+z)^4 \rho_{R,0} \ .
    \end{aligned} 
\end{equation}
We can also convert the differential equation for $\rho_{\Lambda}$ in time to a differential equation of $\rho_{\Lambda}$ in redshift:
\begin{equation} \label{eq:rhoLOfZ}
    \begin{aligned} 
 \frac{d\rho_{\Lambda}}{dt} = \frac{dz}{dt}\frac{d\rho_{\Lambda}}{dz}=  -3 H \rho_\Lambda(1+ w_\Lambda ) 
  \Rightarrow \frac{d\rho_{\Lambda}}{dz} =  -3 H \rho_{\Lambda} (1+ w_\Lambda ) \frac{dt}{dz} \ .
    \end{aligned} 
\end{equation}

We thus arrive at the set of 3 ODEs that we use in Section \ref{sec:altModels}:
\begin{equation} \label{eq:fromWODEs}
  \begin{aligned}
      \frac{d\tau'}{dz} &= \frac{1}{H'} & \tau'(z=0)  = 0\ ,\\
      \frac{d\Omega_{\Lambda}}{dz} &= 3 \Omega_{\Lambda} \frac{1}{1+z} (1 + w_{\Lambda})  &  \Omega_{\Lambda}(z = 0) = \Omega_{\Lambda,0} \ , \\
      \frac{d \ d_L'}{dz} &= \frac{d_L'}{1+z} + \frac{1+z}{H'}  & d_L'(z = 0)  = 0 \ ,
  \end{aligned}
\end{equation}
where 
\begin{equation}
\begin{aligned} \label{eq:justWExtras}
H'  \equiv \frac{H}{H_0} &= \sqrt{\Omega_{m,0} (1+z) ^ 3 + \Omega_{r,0} (1+z) ^ 4 + \Omega_{\Lambda}} \ , \\
w_{\Lambda} &= w_{\Lambda}(\tau' H_0, z)  \ , \\
\tau' & \equiv \tau H_0 \ = \tau h_{100} 100 \textrm{km s}^{-1} \textrm{ Mpc} ^ {-1}  \ ,\\
d_L' & \equiv d_L H_0 /c \ . 
\end{aligned}
\end{equation}
Once $w_{\Lambda} (\tau, z)$ is specified, the ODEs of Equation \ref{eq:fromWODEs} can be numerically solved to determine the values of $d_L$ predicted by this ACM at each measured $\tau h_{100}$ in the Pantheon data set.  

\section{ACMS WITH AN OSCILLATORY DARK ENERGY DENSITY } \label{app:fromDEODEs}
Here, we elaborate on the derivation of the ODEs corresponding to an ACM characterized by an alternate DE energy density that evolves according to,  $\rho_{ \textrm{DM, can}}(z) \rightarrow X(\tau,z) \rho_{ \textrm{DM, can}}(z)$.   
As in Appendix \ref{app:fromWODEs}, our goal is to derive a set of coupled ODEs that can be numerically solved to determine the evolution of cosmic parameters, given an expression for $X(\tau,z)$.  

The ODEs of Equations \ref{eq:tOfZODE} and \ref{eq:tauOfZODE} follow directly from parameter definitions and are thus unchanged.  
We must modify $H(z)$ according to the new DE scaling function: 
\begin{equation} 
    \begin{aligned} 
\frac{H^2(\tau, z)}{H_0^2} 
= ( (1 + z) ^ 3 \Omega_{m,0} + (1 + z) ^ 4 \Omega_{r,0} +  X (\tau, z)  \Omega_{\Lambda,0} )  \ . 
    \end{aligned} 
\end{equation} 

The relevant differential equations for determining the evolution of the SNe Ia luminosity distances in a universe with a scaled DE energy density are then 
\begin{equation} \label{eq:fromDEODEs}
  \begin{aligned}
      \frac{d\tau'}{dz} &= \frac{1}{H'} & \tau'(z=0)  = 0\ ,\\
      \frac{d \ d_L'}{dz} &=   \frac{d_L'}{1+z} + \frac{1+z}{H' }  & d_L'(z = 0)  = 0 \ ,
  \end{aligned}
\end{equation}
where
\begin{equation}
\begin{aligned} \label{eq:justGExtras}
H'  \equiv \frac{H}{H_0}  &=  \sqrt{(1+z) ^ 3 \Omega_{m,0} G'  + (1+z) ^ 4 \Omega_{r,0} G'  + X (\tau, z) \Omega_{\Lambda}} \ , \\
\tau' & \equiv \tau H_0 \ = \tau h_{100} 100 \textrm{km s}^{-1} \textrm{ Mpc} ^ {-1}  \ ,\\
d_L' & \equiv d_L H_0 /c \ .
\end{aligned}
\end{equation}
The ODEs of Equation \ref{eq:fromDEODEs} can be numerically solved once $X(\tau, z)$ is specified.  

\section{ACMS WITH AN EVOLVING GRAVITATIONAL ACCELERATION} \label{app:fromGODEs}
Here, we elaborate on the derivation of the ODEs corresponding to an ACM characterized by an evolving gravitational `constant'.  As in Appendices \ref{app:fromWODEs} and \ref{app:fromDEODEs}, we seek to write down a system of coupled ODEs in $z$ that can be numerically solved for $d_L$. 

The promotion of gravity to a dynamic field, $\phi$, is a popular method for producing a gravitational acceleration that evolves in cosmic time.  Following \cite{Barrow97, Clifton05}, we use the following Equations to describe the cosmic evolution of such a field: 
\begin{equation} \label{eq:GinCosmo1}
H ^ 2 + H \frac{d\phi}{dt} \frac{1}{\phi} - \frac{\omega(\phi)}{6}  \Big (  \frac{d\phi}{dt} \frac{1}{\phi} \Big ) ^2  =  \frac{1}{\phi} \frac{8 \pi}{3c ^ 2} (\rho_M + \rho_R + \rho_\Lambda) \ ,
\end{equation}
\begin{equation}\label{eq:GinCosmo2}
\frac{d \phi^2}{dt^2} + \Big [ 3 H + \frac{d \omega}{dt} \frac{1}{2 \omega(\phi) + 3} \Big ] \frac{d\phi}{dt}  = \frac{3}{2 \omega(\phi) + 3} \frac{8 \pi}{3 c^ 2}  (\rho_M - 3 P_M +  \rho_R - 3 P_R + \rho_\Lambda - 3 P_\Lambda) \ ,
\end{equation}
\begin{equation}
  \begin{aligned}\label{eq:GinCosmo3}
\frac{dH }{dt} + & H ^2 +  \frac{\omega(\phi)}{3} \Big (  \frac{d\phi}{dt} \frac{1}{\phi} \Big ) ^2 - H   \frac{d\phi}{dt} \frac{1}{\phi} \\ & = - \frac{8 \pi}{3 c^ 2}  \frac{\omega(\phi) ( \rho_M + 3P_M + \rho_R + 3P_R + \rho_\Lambda + 3P_\Lambda) + 3 ( \rho_M +  \rho_R +  \rho_\Lambda)}{\phi(2 \omega(\phi) + 3) }  + \frac{1}{2} \frac{d \omega(\phi)}{dt} \frac{1}{2 \omega(\phi) + 3} \frac{d\phi}{dt} \frac{1}{\phi} \ ,
  \end{aligned} 
\end{equation} 
where $\omega(\phi)$ describes the coupling of the field and where we have assumed that the universe is flat. 

Using Equations \ref{eq:rhoMOfZ}-\ref{eq:rhoLOfZ}, defining $\theta \equiv d \phi /dt$, and noting that $d/dz = -1/((1+z) H) d/dt$, we can rewrite Equations \ref{eq:GinCosmo2} and \ref{eq:GinCosmo3} as a set of coupled, first order ODEs: 
\begin{equation}
\frac{d \phi'}{dz} = -\frac{1}{1+z} \frac{1}{H'} \theta' \ ,
\end{equation}
\begin{equation}
\frac{d \theta'}{dz} = \frac{1}{1+z} \frac{\theta'}{H'} \Big [3 H' + \frac{d\omega(\phi')}{d \phi'} \frac{\theta'}{2 \omega(\phi') + 3} \Big ] - \frac{1}{1+z} \frac{1}{H} \frac{3(\Omega_{M, 0} (1+z) ^ 3 + 4 \Omega_{\Lambda, 0})}{2 \omega(\phi ') + 3} \ ,
\end{equation}
\begin{equation}
  \begin{aligned}
\frac{d H'}{dz} =  \frac{H'}{1+z} & + \frac{\omega(\phi')}{3(1+z)H'} \Big ( \frac{\theta'}{\phi'} \Big )^2 - \frac{\theta'}{(1+z) \phi'} - \frac{1}{2} \frac{1}{(1+z)H'} \frac{d \omega(\phi')}{d \phi'} \frac{\theta'^2}{(2 \omega(\phi') + 3)\phi'} \\ 
&+ \frac{\omega(\phi') ( \Omega_{M,0} (1+z) ^ 3 + 2 \Omega_{R,0} (1+z) ^ 4 -2 \Omega_{\Lambda,0}) + 3 ( \Omega_{M,0} (1+z) ^ 3 + \Omega_{R,0} (1+z) ^ 4 + \Omega_{\Lambda,0})}{(1+z) H' \phi' (2 \omega(\phi') + 3) }  \ ,
  \end{aligned}
\end{equation}
where $\phi' \equiv \phi  G_0$, $t' \equiv t H_0$, $H' \equiv H/H_0$, and $\theta' = \theta G_0/H_0$.  We use the additional constraint of Equation \ref{eq:GinCosmo1} to determine the initial value of $\theta'$ (see Equation \ref{eq:fromGODEs}). 

Variation in the strength of gravity would alter the absolute luminosity of SNe Ia.  Following the power law relation of \cite{Gaztanaga01}, we correct the absolute SNe Ia luminosity:
\begin{equation}\label{eq:LofG}
\frac{L_t}{L} = \Big( \frac{G}{G_0} \Big) ^ {\alpha} \ ,
\end{equation}
where $L$ is the standard absolute SNe Ia luminosity, $L_t$ is the true absolute SNe Ia luminosity, and $\alpha$ is a parameter specifying the strength of the dependence of the true SN luminosity on the local value of $G$.  \cite{Gaztanaga01} claim that $\alpha \simeq -3/2$ and we adopt that value here.  However, we do note that alternate relations between $L$ and $G$ have been considered by \cite{Wright18}.  

Because the luminosity distance of observed SNe Ia, $d_L$, is defined according to the absolute luminosity of SNe Ia in our local universe (Equation \ref{eq:dLDef}), an evolving gravitational acceleration would alter the measurements of $d_L$:
\begin{equation}
    \begin{aligned} 
d_L  \equiv \sqrt{\frac{L}{4 \pi f}} 
        = \sqrt{\frac{L}{4\pi}}  \sqrt{\frac{4 \pi}{L_t}} c(1+z)  \displaystyle \int_0^z dz' \frac{1}{\sqrt{H(z')}} 
        = \Big ( \frac{G_0}{G} \Big ) ^ {\alpha/2} c (1 + z)\displaystyle \int_0^z dz' \frac{1}{\sqrt{H(z')}} \ ,
     \end{aligned} 
\end{equation}
where we have noted that $f$ is dependent on $L_t$, but $d_L$ is defined using $L$.  The ODE dictating the evolution of $d_L$ with $z$ thus picks up an additional term proportional to the variation in $G(\tau,z)$:
\begin{equation}\label{eq:dLOfZODEFromG}
    \begin{aligned} 
\frac{d\ d_L}{dz} &= \Big ( \frac{G_0}{G} \Big ) ^ {\alpha/2} \frac{c(1+z)}{H(z)} 
 - \Big ( \frac{G_0}{G} \Big ) ^ {\alpha/2} \frac{\alpha} {2} \frac{1}{G} \frac{dG}{dz} c(1+z) \int_0^z dz' \frac{1}{H(z')} 
 + \Big ( \frac{G_0}{G} \Big ) ^ {\alpha/2} c \int_0^z dz' \frac{1}{H(z')} \\
 &= \frac{d_L}{(1+z)} + \Big ( \frac{G_0}{G} \Big ) ^ {\alpha/2} \frac{c(1+z)}{H(z)} -\frac{\alpha}{2} \frac{d_L}{ G}\frac{dG}{dz}   \ .
    \end{aligned} 
\end{equation}

We thus arrive at the set of four ODEs used in Section \ref{sec:altModels}:
\begin{equation} \label{eq:fromGODEs}
  \begin{aligned}
\frac{d \phi'}{dz} & = -\frac{1}{1+z} \frac{1}{H'} \theta'  & \phi'(z=0) = \phi'_0 \geq \frac{2 \omega(\phi'_0)}{3 + 2 \omega(\phi'_0)} \ , \\ 
\frac{d \theta'}{dz} &= \frac{1}{1+z} \frac{\theta'}{H'} \Big [3 H' + \frac{d\omega(\phi')}{d \phi'} \frac{\theta'}{2 \omega(\phi') + 3} \Big ] \\
& \ \ \ \ \ \ \ \ \ \ \ \ \ \ - \frac{1}{1+z} \frac{1}{H} \frac{3(\Omega_{M, 0} (1+z) ^ 3 + 4 \Omega_{\Lambda, 0})}{2 \omega(\phi ') + 3}  & \theta'(z=0) = \frac{3 \phi'_0}{\omega(\phi'_0)} + \sqrt{ \frac{3 \phi'_0}{\omega(\phi'_0)} \Big (\frac{3 \phi'_0}{\omega(\phi'_0)} + 2 \phi'_0 - 2 \Big ) } \ , \\
\frac{d H'}{dz} & =  \frac{H'}{1+z}  + \frac{\omega(\phi')}{3(1+z)H'} \Big ( \frac{\theta'}{\phi'} \Big )^2 \\
&\ \ \ - \frac{\theta'}{(1+z) \phi'} - \frac{1}{2} \frac{1}{(1+z)H'} \frac{d \omega(\phi')}{d \phi'} \frac{\theta'^2}{(2 \omega(\phi') + 3)\phi'} \\ 
&\ \ \ + \frac{\omega(\phi') ( \Omega_{M,0} (1+z) ^ 3 + 2 \Omega_{R,0} (1+z) ^ 4 -2 \Omega_{\Lambda,0})}{(1+z) H' \phi' (2 \omega(\phi') + 3) } \\ 
& \ \ \ \ + \frac{3 ( \Omega_{M,0} (1+z) ^ 3 + \Omega_{R,0} (1+z) ^ 4 + \Omega_{\Lambda,0})}{(1+z) H' \phi' (2 \omega(\phi') + 3) }  & H'(z = 0) = 1 \ , \\
      \frac{d \ d_L'}{dz} &=  \frac{d_L'}{1+z} +  \Big ( \frac{1}{G'} \Big ) ^ {\alpha/2} \frac{1+z}{H'\ } - \frac{\alpha}{2} \frac{d_L'}{G'}   \frac{dG'}{dz} 
                                     & d_L'(z = 0)  = 0 \ ,
  \end{aligned}
\end{equation}
where 
\begin{equation}
\begin{aligned} \label{eq:justGExtras}
\phi' & \equiv \phi G_0 \ , \\
G'  & \equiv G / G_0= \frac{1}{\phi'}  \frac{4+2 \omega(\phi')}{3+2\omega(\phi')} \ , \\
d_L' & \equiv d_L H_0 /c \ , \\
\theta' &\equiv \theta G_0 / H_0 \ , \\
H' & \equiv H/H_0 \ ,
\end{aligned}
\end{equation}
where $G_0$ is Newton's constant and where $\phi'_0$, the value of the $\phi'$ field today, is determined by inverting the expression for $G'$ in Equation \ref{eq:justGExtras} with $G'=1$.  The relationship between $G$ and $\phi$ is derived in \cite{Nordtvedt70}. 

The ODEs of Equation \ref{eq:fromGODEs} describe how $d_L$, and thus $\mu$, would evolve with $z$ if gravity derived from a scalar field that evolves according to the some specified coupling, $\omega(\phi)$.  

\section{CONSTRAINTS ON THE CONTEMPORARY VARIATION OF $G$} \label{app:lunarGConstraints}
In constraining ACMs characterized by an evolving scalar field of gravity, we must note that there are several contemporary constraints on the rate of change of $G$.  The strongest such constraint comes from the Lunar Ranging Experiment \citep{Muller07}, which reports the following measurements on the contemporary rate of change of $G$:
\begin{equation} \label{eq:myLunarConstraints} 
    \begin{aligned}
    \frac{dG/dt}{G} \Big | _{t = t_0}      &= \frac{1}{G_0} \Big ( \frac{dz}{dt} \frac{dG}{dz} \Big )\Big |_{z = t_0} 
                                                 = \frac{1}{G_0} \Big ( -(1+z) H \frac{dG}{dz}\Big )\Big |_{z = 0} 
                                                 = -\frac{H_0}{G_0} \Big ( \frac{dG}{dz}  \Big )\Big |_{z = 0} 
                                                 = -(2\pm7) \times 10 ^{-13}\textrm{ yr}^{-1}\ ,\\
    \frac{d^2G/dt^2}{G}\Big | _{t=t_0} &=  \frac{1}{G_0}  \Big [-(1+z)H \frac{d}{dz} \Big ( -(1+z) H \frac{dG}{dz}\Big ) \Big ] \Big |_{z = 0} \\ 
                                               & = \frac{H_0^2}{G_0}  \Big (\frac{dG}{dz} + \frac{dH'}{dz} \frac{dG}{dz} +\frac{d^2G}{dz^2 }  \Big ) \Big |_{z = 0}                                             = -(4\pm5) \times 10 ^{-15}\textrm{ yr}^{-2} \ ,
    \end{aligned}
\end{equation} 
where we have converted from derivatives in $t$ to derivatives in $z$, noted that $dz/dt= -(1+z)H$, and applied an overall factor of $-1$ because we define $t$ looking back in time while \cite{Muller07} define $t$ moving forward.  

Equations \ref{eq:fromGODEs} and \ref{eq:justGExtras} can be combined with the constraints in Equation \ref{eq:myLunarConstraints} to determine if a particular choice of $\omega(\phi)$ is consistent with our contemporary knowledge of the strength and invariance of gravity.  We show the $5\sigma$ limits of these constraints for our choice of $\omega(\phi)$ (Equation \ref{eq:GOfTDef}) as dark shading in Figure \ref{fig:fromGFourierLims}. 

\bibliography{refs}

\end{document}